\documentclass[twocolumn, twocolappendix]{aastex63}
	
\usepackage[encapsulated]{CJK}
\usepackage{ucs}
\usepackage[utf8x]{inputenc}
\usepackage{url}
\newcommand{\cntext}[1]{\begin{CJK}{UTF8}{bsmi}#1\end{CJK}}

\usepackage[italicdiff]{physics} 
\usepackage{amsmath}

\usepackage{graphicx}

\usepackage{braket}

\usepackage{xcolor}

\newcommand{\m}[1]{\underline{\underline{#1}}}

\newcommand{\coone}{\mbox{CO(1--0)}}
\newcommand{\cotwo}{\mbox{CO(2--1)}}

\newcommand{\cii}{\textsc{C\,ii}}
\newcommand{\hi}{\textsc{H\,i}}
\newcommand{\hii}{\textsc{H\,ii}}

\shorttitle{Modeling $X_{\rm CO}$ \& $R_{21}$ in a Multiphase ISM}
\shortauthors{Hu et al.}

\begin{document}

\defcitealias{HSvD21}{HSvD21}


\title{Dependence of $X_{\rm CO}$ on metallicity, intensity, and spatial scale
	in a self-regulated interstellar medium}
\author[0000-0002-9235-3529]{Chia-Yu Hu (\cntext{胡家瑜})}
\affiliation{Max-Planck-Institut f\"{u}r Extraterrestrische Physik, Giessenbachstrasse 1, D-85748 Garching, Germany}
\author{Andreas Schruba}
\affiliation{Max-Planck-Institut f\"{u}r Extraterrestrische Physik, Giessenbachstrasse 1, D-85748 Garching, Germany}
\author{Amiel Sternberg}
\affiliation{School of Physics \& Astronomy, Tel Aviv University, Ramat Aviv 69978, Israel}
\affiliation{Center for Computational Astrophysics, Flatiron Institute, 162 5th Ave, New York, NY 10010, USA}
\affiliation{Max-Planck-Institut f\"{u}r Extraterrestrische Physik, Giessenbachstrasse 1, D-85748 Garching, Germany}
\author{Ewine F. van Dishoeck}
\affiliation{Max-Planck-Institut f\"{u}r Extraterrestrische Physik, Giessenbachstrasse 1, D-85748 Garching, Germany}
\affiliation{Leiden Observatory, Leiden University, P.O. Box 9513, NL-2300 RA Leiden, the Netherlands}
\correspondingauthor{Chia-Yu Hu}
\email{cyhu.astro@gmail.com}

\begin{abstract}

We study the \coone-to-H$_2$ conversion factor ($X_{\rm CO}$) and the line ratio of \cotwo-to-\coone\ ($R_{21}$)
across a wide range of metallicity ($0.1 \leq Z/Z_\odot \leq 3$)
in high-resolution (${\sim}0.2$~pc) hydrodynamical simulations
of a self-regulated multiphase interstellar medium.
We construct synthetic CO emission maps via radiative transfer
and systematically vary the ``observational'' beam size to quantify the scale dependence.
We find that
the kpc-scale
$X_{\rm CO}$
can be over-estimated at low~$Z$
if assuming steady-state chemistry
or assuming that the star-forming gas is H$_2$-dominated.
On parsec scales,
$X_{\rm CO}$ varies by orders of magnitude from place to place, 
primarily driven by 
the transition from atomic carbon to CO.
The pc-scale $X_{\rm CO}$
drops to the Milky Way value of $2\times 10^{20}\ {\rm cm^{-2}~(K~km~s^{-1})^{-1}}$ once dust shielding becomes effective, independent of $Z$.
The CO lines become increasingly optically thin at lower $Z$,
leading to a higher $R_{21}$.
Most cloud area is filled by diffuse gas with high $X_{\rm CO}$ and low $R_{21}$,
while most CO emission originates from dense gas with low $X_{\rm CO}$ and high $R_{21}$.
Adopting a constant $X_{\rm CO}$ strongly over- (under-)estimates H$_2$ in dense (diffuse) gas.
The line intensity 
negatively (positively) correlates with $X_{\rm CO}$ ($R_{21}$)
as it is a proxy of column density (volume density).
On large scales, $X_{\rm CO}$ and $R_{21}$ are dictated by beam averaging,
and they are naturally biased towards values in dense gas.
Our predicted $X_{\rm CO}$ is a multivariate function of $Z$, line intensity, and beam size,
which can be used to more accurately infer the H$_2$ mass.

\end{abstract}
\keywords{Interstellar medium (847); Astrochemistry (75); Hydrodynamical
simulations (767)}

\section{Introduction} \label{sec:intro}

Molecular clouds are the site for star formation.
Characterizing how they form from, and interact with, 
the diffuse interstellar medium (ISM)
is crucial for our understanding of galaxy evolution \citep{2003ARA&A..41...57L, 2007ARA&A..45..565M, 2020ARA&A..58..157T}.
However,
molecular hydrogen (H$_2$),
the main component of molecular clouds,
does not emit radiation under typical conditions.
In contrast,
carbon monoxide (CO),
the second most abundant molecular species,
is a very efficient emitter at low temperatures
and therefore is the most commonly used tracer for H$_2$.
The CO-to-H$_2$ conversion factor
\begin{equation}
	X_{\rm CO} \equiv \frac{N_{\rm H_2}}{W_{10}}~,
\end{equation}
is used to convert the observed intensity of the \coone\ line intensity ($W_{10}$)
to the H$_2$ column density ($N_{\rm H_2}$).
The typical value found in the Milky Way is
\begin{equation}\label{eq:XCO_MW}
X_{\rm CO,MW} = 2\times 10^{20}\ {\rm cm^{-2}~(K~km~s^{-1})^{-1}}
\end{equation}
within $\pm 30\%$ uncertainty (see \citealp{2013ARA&A..51..207B} for a detailed review).

\begin{figure*}
	\centering
	\includegraphics[trim=0cm 0cm 0cm 1cm,clip, width=0.7\linewidth]{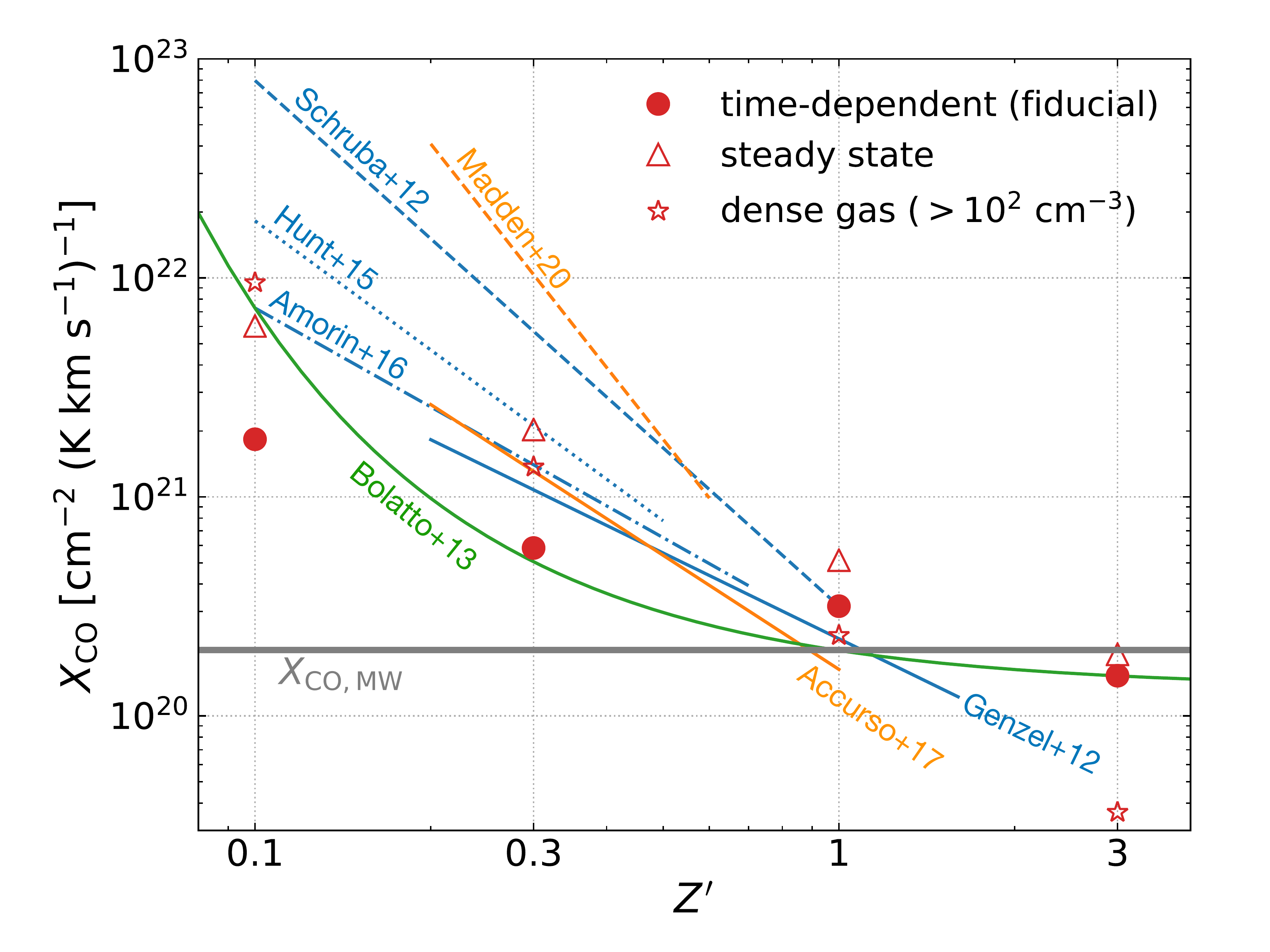}
	\caption{%
	    CO-to-H$_2$ conversion factor ($X_{\rm CO}$) as a function of normalized metallicity ($Z^\prime \equiv Z / Z_\odot$). The colored lines are from different observational studies as follows (showing only the dynamical range of $Z^\prime$ in each observation). \citet{2012ApJ...746...69G}: high-redshift ($z>1$) star-forming galaxies (inverse KS~method); \citet{2012AJ....143..138S}: dwarf galaxies (inverse KS~method); \citet{2015A&A...583A.114H}: dwarf galaxies (inverse KS~method); \citet{2016A&A...588A..23A}: blue compact dwarf galaxies (inverse KS~method); \citet{2017MNRAS.470.4750A}: local star-forming galaxies (spectral synthesis method); \citet{2020A&A...643A.141M}: dwarf galaxies (spectral synthesis method). The green solid curve shows the theoretical prescription from \citet{2013ARA&A..51..207B}. The horizontal grey line shows the standard Milky Way value (see Eq.~\eqref{eq:XCO_MW}).
	    The red symbols show the time-averaged global (1~kpc$^2$) $X_{\rm CO}$ in our simulations (see Section \ref{sec:global}).
		Red filled circles: our fiducial, time-dependent model; red non-filled triangles: the steady-state model; red non-filled stars: the ``dense gas'' conversion factor $X_{100} \equiv N_{\rm H}(n > 100~{\rm cm^{-3}}) / W_{10}$. 
		{Our time-dependent model can be described by Eq.~(\ref{eq:fit1}). }
		At low $Z^\prime$, $X_{\rm CO}$ can be overestimated both by the spectral synthesis method (by assuming steady-state chemistry) and by the inverse KS~method (by assuming the star-forming gas is fully molecular).
	}
	\label{fig:intro_global_Z_XCO}
\end{figure*}

However,
CO is not a perfect tracer for H$_2$.
Even at solar metallicity,
CO is photodissociated deeper into the cloud by the far-ultra\-violet (UV) radiation 
compared to H$_2$ \citep{1988ApJ...334..771V, 1995ApJS...99..565S},
leading to a regime where H$_2$ gas is deficient in CO,
commonly referred to as the ``CO-dark'' molecular gas \citep{2010ApJ...716.1191W}.
Furthermore,
at low metallicities,
the far-UV radiation 
progressively
photodissociates more CO 
but only mildly so the H$_2$ \citep{1997ApJ...483..200M, 1998ApJ...498..735P, 1999ApJ...513..275B, 2010ApJ...716.1191W},
which expands the range of column densities where gas is CO-dark and leads to a higher $X_{\rm CO}$,
making CO a poor tracer of H$_2$.
Indeed,
observations of nearby low-metal\-licity dwarf galaxies
have found that the CO emission is extremely weak (often undetectable),
and
the ratio of the total star formation rate (SFR) to the CO luminosity
is much higher than what is found in typical spiral galaxies \citep{2012AJ....143..138S, 2014A&A...564A.121C}.
If we assume that the ratio of total SFR to H$_2$ mass is insensitive to metallicity,
this implies an elevated $X_{\rm CO}$ at low metallicities.
Therefore,
theoretical expectations and observations both suggest that
$X_{\rm CO}$ increases at lower metallicity.

Observational measurements of $X_{\rm CO}$ are challenging
as they require not only a successful CO detection (which is difficult at low metallicities)
but also an independent method to derive the H$_2$ mass.
The existing techniques include the following:
(1) The inverse Kennicutt--Schmidt (KS) method
\citep{2012ApJ...746...69G, 2012AJ....143..138S, 2015A&A...583A.114H, 2016A&A...588A..23A}
measures the SFR
which is converted to the associated H$_2$ mass 
via an adopted correlation between the two quantities (i.e., the KS relation).
(2) The dust-based method
\citep{2011ApJ...737...12L, 2013Natur.495..487E, 2015ApJ...804L..11S, 2015MNRAS.450.2708L, 2017ApJ...835..278S}
uses infrared (IR) continuum measurements to derive the dust mass,
which is converted to the total gas mass with an assumed dust-to-gas ratio (DGR).
The H$_2$ mass is then derived by subtracting the atomic hydrogen (\hi) mass (obtained through 21~cm emission) 
from the total gas mass.
(3) The virial method \citep{2008ApJ...686..948B, 2015Natur.525..218R, 2017ApJ...835..278S}
measures the CO line width and the cloud size to derive the H$_2$ mass 
assuming that clouds are in virial equilibrium.
(4) The spectral synthesis method \citep{2017MNRAS.470.4750A, 2020A&A...643A.141M}
measures the fine structure ``metal'' lines (e.g.~[\cii] 158~$\mu$m) in addition to the \coone\ line,
and uses chemistry and spectral synthesis codes with assumed cloud geometry
to constrain the physical parameters of the ISM and derive the H$_2$ mass.
Each of these methods relies on different assumptions 
that are mostly calibrated at solar metallicity,
and their validity at low metallicities remains unclear.
Therefore,
it is perhaps not too surprising
that 
there are still significant discrepancies on the $Z{-}X_{\rm CO}$ relation
derived from different observations (see Fig.~\ref{fig:intro_global_Z_XCO}).
On the other hand,
the discrepancies may also indicate secondary dependencies.
For example,
the $Z{-}X_{\rm CO}$ relation appears to be steeper 
in dwarf galaxies than in more massive, star-forming galaxies
in Fig.~\ref{fig:intro_global_Z_XCO}.

Recently,
high-resolution observations with the Atacama Large Millimeter/submillimeter Array (ALMA)
have successfully detected CO at $Z^\prime \equiv Z / Z_\odot \sim 0.1$
and shown that CO typically originates from compact regions of a few parsecs
\citep{2015Natur.525..218R, 2015ApJ...804L..11S, 2017ApJ...835..278S}.
More importantly,
the $X_{\rm CO}$ derived from the virial method
is found to be comparable to the Milky Way value
and 
is significantly lower than the $X_{\rm CO}$ of the same targets
derived from the dust-based method.
This suggests a dependence on spatial scale (i.e., the beam size) as
the high-resolution virial method measures $X_{\rm CO}$ in the compact (pc-scale), CO-bright cores
while the dust-based method, 
which typically has a much larger beam size ($\gtrsim 100$~pc),
measures $X_{\rm CO}$ averaging over 
entire molecular clouds (or associations), including
both CO-bright and CO-dark gas.
However,
the fundamental small-scale distribution of $X_{\rm CO}$, 
which dictates $X_{\rm CO}$ on different (larger) scales,
is still poorly understood.
The main goal of this paper is to shed light on this subject 
in a systematic way.

Meanwhile,
the \cotwo\ line has become more widely used as an alternative to \coone\ in recent years.
For example,
the recent high-resolution survey of nearby star-forming galaxies,
PHANGS--ALMA \citep{2021ApJS..257...43L},
chose \cotwo\ over \coone\ for better sensitivity.
In this case,
inferring H$_2$ requires a two-step process:
the observed \cotwo\ line intensity ($W_{21}$)
is first converted into $W_{10}$ with an adopted line ratio
\begin{equation}
	R_{21} \equiv \frac{W_{21}}{W_{10}}~,
\end{equation}
and then converted into $N_{\rm H_2}$ with an adopted $X_{\rm CO}$.
Understanding how $R_{21}$ varies with ISM properties is therefore
important for \cotwo\ to be used as an alternative H$_2$ tracer \citep{2021MNRAS.504.3221D}.
Ultimately,
for high-redshift galaxies,
a similar method may be needed for CO lines of higher rotational levels.

Hydrodynamical simulations have been the theoretical frontiers of forward modeling
the ISM structure, chemistry, and the observed line intensities.
However, earlier studies have fundamental limitations in their setup.
On small scales,
cloud simulations
\citep[e.g.,][]{2011MNRAS.412..337G, 2011MNRAS.412.1686S, 
	2011MNRAS.415.3253S, 2012MNRAS.426..377G, 2017ApJ...839...90B,
	2017MNRAS.465.2277P, 2018MNRAS.475.1508P, 2021MNRAS.502.2701B}
provide detailed information with 
very high resolution (${\lesssim}0.1$~pc)
but rely on unrealistic boundary conditions (e.g., isolated clouds) 
and artificial forces for turbulence driving.
In addition,
they do not provide information on the ISM on kpc scales
that is typical for extragalactic observations.
On the other hand,
large-scale galaxy simulations
with resolution ${\gtrsim}50$~pc
\citep[e.g.,][]{2012MNRAS.421.3127N, 2012ApJ...747..124F}
cannot resolve the structures of molecular clouds
and therefore rely on over-simplified assumptions on the sub-grid gas distribution.
Significantly better resolutions have been achieved
in isolated Milky Way-like galaxy simulations
\citep[e.g.,][]{2015MNRAS.447.2144D, 2016MNRAS.458..270R},
but they were still insufficient to properly resolve the ISM.

Recently,
it has become feasible to simulate a kpc-scale ISM patch 
and follow star formation and stellar feedback self-consistently,
establishing a more realistic, self-regulated system 
with pc-scale resolution
such that the ISM structure and stellar feedback are resolved
without resorting to sub-grid models \citep{2017MNRAS.466.1903G, 2017MNRAS.472.4797S, 2018ApJ...858...16G, 2020MNRAS.492.1465S, 2020MNRAS.492.1594S}.
However,
most of these studies focus on solar-metallicity, solar-neighborhood conditions, 
and the only exception \citep{2020ApJ...903..142G} studied $X_{\rm CO}$ 
with a limited range of $0.5 \leq Z^\prime \leq 2$,
which still does not cover the ``low-metal\-licity'' regime.

In \citet{HSvD21} (hereafter \citetalias{HSvD21}),
we presented a suite of hydrodynamical simulations 
of a (1~kpc)$^2$ ISM patch
with an unprecedented dynamical range spatially and temporally,
running for 500~Myr and
reaching sub-parsec (${\sim}0.2$~pc) spatial resolution
(which has previously only been achieved with the ``zoom-in'' techniques and was limited to a few Myrs, e.g., \citealp{2020MNRAS.492.1465S, 2020MNRAS.492.1594S}).
The long simulated time leads to a large sample of clouds at different evolutionary stages,
while the sub-parsec resolution is crucial for resolving
the compact CO-bright cores at low metallicities. 
Furthermore,
the simulations cover a wide range of metallicities ($0.1 \leq Z^\prime \leq 3$),
probing the low-metal\-licity regime for the first time.
The numerical framework
self-consistently includes
gravity, hydrodynamics, 
time-dependent cooling and H$_2$ chemistry, 
radiation shielding, 
individual star formation, 
and 
stellar feedback from supernovae and photoionization.
To simultaneously resolve the ISM
and model the chemistry accurately,
a hybrid approach is introduced
where H$_2$ is followed on-the-fly
while the other chemical species, including CO, are
modeled in post-processing by a more detailed network.

In this paper,
we construct synthetic emission maps of \coone\ and \cotwo\ from the simulations in \citetalias{HSvD21}
with radiative transfer calculations.
Our goal 
is to investigate the distributions of $X_{\rm CO}$ and $R_{21}$ 
from pc to kpc scales at different metallicities
and how they affect $X_{\rm CO}$ and $R_{21}$ on systematically larger scales. 
Our paper is organized as follows.
Section~\ref{sec:method}
describes our numerical setup for the radiative transfer calculation on an adaptive mesh.
Section~\ref{sec:results} is an overview of our simulation results,
including our global $X_{\rm CO}$ compared against observations.
Section~\ref{sec:mol_physics} is a detailed analysis on the small-scale properties of CO excitation and optical depth.
Section~\ref{sec:obs_impl} shows our predicted $X_{\rm CO}$ and $R_{21}$ on different spatial scales
where we systematically increase the beam size from 2~pc to 1~kpc.
Section~\ref{sec:summary} summarizes our work.

\section{Numerical methods} \label{sec:method}

\subsection{Simulations}

Our simulations are presented in detail in \citetalias{HSvD21},
which we briefly summarize as follows.
The setup is an ISM patch with conditions similar to the solar neighborhood.
The box size is 1~kpc along the $x$- and $y$-axes with periodic boundary conditions 
and 10~kpc along the $z$-axis with outflow boundary conditions.
The origin is defined at the box center and $z = 0$ represents the mid-plane of the disk.
The simulations were conducted using the
public version of {\sc Gizmo} \citep{2015MNRAS.450...53H},
a multi-solver code featuring the meshless Godunov method \citep{2011MNRAS.414..129G} 
built on the TreeSPH code {\sc Gadget-3} \citep{2005MNRAS.364.1105S}.
Gravity is solved by the ``treecode'' method \citep{1986Natur.324..446B} 
while
hydrodynamics is solved by 
the meshless finite-mass (MFM) method \citep{2015MNRAS.450...53H}.
Time-dependent cooling and H$_2$ chemistry are included based on \citet{2007ApJS..169..239G} and \citet{2012MNRAS.421..116G},
with a {\sc HealPix} \citep{2011ascl.soft07018G}-based treatment for radiation shielding
similar to \citet{2012MNRAS.420..745C}.
Star formation is based on the commonly used, stochastic ``Schmidt law'' recipe 
with a star formation efficiency of 50\%.
The stellar masses are stochastically sampled from the stellar initial mass function of  \citet{2001MNRAS.322..231K}
which determine the lifetime of massive stars \citep{2012A&A...537A.146E}
and the luminosity of ionizing radiation \citep{1997A&AS..125..229L, 1998A&AS..130...65L}.
Supernova feedback is purely thermal.
Photoionization follows the method of \citet{2017MNRAS.471.2151H} 
that can properly account for overlapping \hii\ regions.
The far-UV radiation field and cosmic-ray ionization rate are 
both spatially uniform but time-dependent,
scaled linearly with the total star formation rate.
The metallicity is assumed to be constant both spatially and temporally.

Four simulations are run with metallicities of $Z^\prime = 3, 1, 0.3, \text{and } 0.1$.
The dust-to-gas mass ratio is 1\% at $Z^\prime = 1$ and
scales linearly with $Z^\prime$.
We first run the $Z^\prime = 1$ model with an artificially less efficient feedback model for 100 Myr in order to mitigate the initial transient phase that tends to blow out the gaseous disk.
This generates a multiphase ISM structure which is then used by each simulation as the initial conditions.
Each simulation is run for 500~Myr.
The mass resolution is $1~{\rm M_\odot}$ per gas particle,
which corresponds to ${\sim}0.2$~pc spatial resolution where the Jeans mass is resolved.

The results are further post-processed with a more detailed chemistry network 
that includes CO chemistry,
using the time-dependent H$_2$ abundances from the simulations.
We assume that an external FUV radiation background (calculated from the SFR)
is attenuated by the effects of dust shielding,
H$_2$ self-shielding, and CO self- and mutual-shielding.
The column densities of dust, H$_2$, and CO relevant for shielding are calculated using a pixel-based approach
integrated up to a radius of 100~pc for each gas cell.
The photodissociation rate coefficients and attenuation coefficients are taken from
\citet{2017A&A...602A.105H}\footnote{\url{https://home.strw.leidenuniv.nl/~ewine/photo/index.html}}.
The time-averaged SFR in all runs are close to the observed value in the solar-neighborhood of $\Sigma_{\rm SFR,0} = 2.4\times 10^{-3}\ {\rm M_\odot\ yr^{-1}\ kpc^{-2}}$ \citep{2009AJ....137..266F} within a factor of 50\%.
Therefore,
the corresponding time-averaged FUV radiation field is close to the Draine field \citep{1978ApJS...36..595D} and the cosmic-ray ionization rate is close to 
$\zeta_{\rm CR} = 10^{-16}\ {\rm s^{-1}}$ \citep{2012ApJ...745...91I, 2015ApJ...800...40I} in all runs.

\subsection{Radiative Transfer on an Adaptive Mesh}\label{sec:AMR}

We post-process our simulations with the publicly available radiative transfer code {\sc RADMC-3D} \citep{2012ascl.soft02015D} to generate the synthetic \coone\ and \cotwo\ emission maps in the face-on view. We describe the details as follows.

Before we perform radiative transfer, we first need to convert the particle-based gas properties in our simulations onto a mesh where {\sc RADMC-3D} calculates the level population and the subsequent radiative transfer. Our simulations resolve the Jeans mass of the gas down to ${\sim}0.2$~pc, which defines the smallest spatial scale. However, if we were to generate a uniform Cartesian mesh that covers the entire simulation domain with a spatial resolution of $0.2$~pc, it would take $5000^3$ cells which is computationally infeasible both in terms of computing time and memory. In addition, a large fraction of the simulation domain is filled with CO-free diffuse gas, making a uniform mesh extremely computationally inefficient. As such, it is highly desirable to adopt an adaptive mesh  where the finest cells are only employed in regions of dense molecular gas.

We generate an adaptive mesh by first building an octree where each leaf contains at most one gas particle. The root node of the octree is centered at $x = 0.5$~kpc, $y = 0.5$~kpc, and $z = 0$, with a size of $1$~kpc on each side, which contains the vast majority of the gas.%
\footnote{%
    Gas at high latitude ($z > 0.5$~kpc) is essentially CO-free and therefore can be excluded without affecting our results.}
This octree not only defines an adaptive mesh where the resolution improves (i.e., cell size decreases) with increasing gas density but also serves as a neighbor finder for interpolation (see below). As the smallest physically meaningful scale in our simulations is around $0.2$~pc, we trim the mesh such that the size of the smallest cell is $h_{\rm min} = 1/2^{13}~\mathrm{kpc} = 0.12$~pc  (i.e., the maximum refinement level is~13). The mesh is then further refined to fulfill the ``$2\,{:}\,1$ balance,'' which avoids a sudden jump of refinement levels and optimizes the calculation of velocity gradients in {\sc RADMC-3D}.

Once the adaptive mesh is generated, we interpolate the particle information onto the mesh with a scheme similar to the MFM method. For any scalar field~$A$, the interpolated value at each cell center $x_c$ is
\begin{equation}
	\bar{A}(x_c) = \frac{ \sum_{j} A_j K(|x_j - x_c|, h_j) }{ \sum_{j} K(|x_j - x_c|, h_j) }~,
\end{equation}
where $x_j$ and $h_j$ are the location and smoothing length of particle~$j$, respectively, $K(|x_j - x_c|, h_j) = \omega(|x_j - x_c| / h_j) / h_j^3$, and $\omega$ is the cubic spline kernel function. The summation is over all particles whose kernel overlaps with $x_c$ (i.e., $|x_j - x_c| \leq h_j$), which is done by a ``scatter''-neighbor search on $x_c$ using the particle octree. We find that this interpolation scheme is significantly more accurate than  the conventional smoothed-particle hydrodynamics (SPH) scheme especially for the gas temperature. The effect of different interpolation schemes is discussed in Appendix~\ref{app:interp}. We interpolate the following quantities onto the adaptive mesh: CO number density ($n_{\rm CO}$), H$_2$ number density ($n_{\rm H_2}$), gas temperature ($T$), velocity ($v$), and velocity dispersion%
\footnote{%
    We first calculate the particle-based $\sigma_v$ as the velocity dispersion within the gas smoothing length and then interpolate it onto the mesh.}
($\sigma_v$).

We use the large-velocity gradient approximation module implemented by \citet{2011MNRAS.412.1686S} in {\sc RADMC-3D} to account for radiation trapping and calculate the level population at each cell. The molecular data are taken from the Leiden Atomic and Molecular database \citep{2005A&A...432..369S}. The collision partner of CO is H$_2$ with an ortho-to-para ratio of~3, and the collisional rate coefficients are taken from \citet{2001JPhB...34.2731F} and \citet{2006A&A...446..367W}. The local line profile follows a Gaussian function
\begin{equation}
	\phi(\nu) = \lambda_0 / (\sqrt{\pi} b) e^{-\Delta v^2/b^2}~,
\end{equation}
where  $\nu$ is the photon frequency, $\lambda_0 \equiv c / \nu_0$ ($c$~is the speed of light and $\nu_0$ is the line-center photon frequency), and $\Delta v \equiv c (\nu-\nu_0) / \nu_0$ is the Doppler velocity offset. The Doppler parameter $b\equiv \sqrt{2 k_\mathrm{B} T / (\mu m_\mathrm{p}) + \sigma_v^2}$ accounts for both thermal and microturbulent line broadening, where $k_\mathrm{B}$ is the Boltzmann constant, $m_\mathrm{p}$~is the proton mass, and $\mu$ is the mean molecular weight.

Once the level populations are known, {\sc RADMC-3D} solves the equation of radiative transfer along the $z$-axis through the adaptive mesh. For each simulation snapshot, we adopt $512 \times 512$ pixels/\linebreak[0]{}rays that cover the entire simulation domain of $1~\mathrm{kpc} \times 1~\mathrm{kpc}$. This means a pixel size of $1000/512 \approx 2$~pc, which is 16~times coarser than our finest cell. It might therefore seem concerning that a substantial fraction of small cells could be missed by the rays. Fortunately, {\sc RADMC-3D} has the capability of ``recursive sub-pixeling,'' where a ray is split into 2-by-2 rays recursively (similar to a quadtree)  whenever it passes through a cell smaller than the pixel size. As our pixels coincide with the octree's hierarchical structure, all cells, including those smaller than $2$~pc, are guaranteed to be properly ``visited'' by the rays and so flux conservation is ensured.  This is of crucial importance especially at low metallicities as CO typically concentrates in small dense cores.

At each pixel, {\sc RADMC-3D} produces a spectrum, i.e., the specific intensity $I_\nu$ as a function of~$\nu$. It is convenient to convert $I_\nu$ into radiation temperature
\begin{equation}
    T_\mathrm{R} \equiv \frac{\lambda_{0}^2}{2 k_\mathrm{B}} I_\nu~.
\end{equation}
The line intensity is defined as the velocity-integrated radiation temperature 
\begin{equation}
    W \equiv \int T_\mathrm{R} \,\mathrm{d}v~.
\end{equation}
which is in units of ${\rm K~km~s^{-1}}$. The numerical integration is done with the Simpson's rule. The velocity coverage of the emission spectra we adopt is $\pm 20~{\rm km~s^{-1}}$, which is wide enough to cover the vertical motion of the gas. The velocity resolution is $0.4~{\rm km~s^{-1}}$, which is sufficient considering the thermal broadening alone leads to $b \approx 1~{\rm km~s^{-1}}$ for $T = 100$~K.
An example of typical spectra is shown in Appendix~\ref{app:spectra}.

For each snapshot, we run {\sc RADMC-3D} following the above-mentioned approach and generate $512 \times 512$ spectra of both the \coone\ and \cotwo\ transitions as well as their corresponding maps of line intensity $W_{10}$ and $W_{21}$. We also obtain the number density of CO in the rotational levels $J = 0, 1, \text{and } 2$ on the adaptive mesh,  denoted as $n_0$, $n_1$, and $n_2$, respectively. We conduct the calculation for 41 snapshots from $100$ to $500$~Myr with a time interval of $10$~Myr. Throughout the paper, all results we show are time-averaged over $400$~Myr, except for the snapshots in Figs.~\ref{fig:co_10_21_maps} and~\ref{fig:XCO_maps_pix}.


\section{Overview of simulation results}\label{sec:results}

\subsection{Global Average Quantities}\label{sec:global}

\begin{figure}
	\centering
	\includegraphics[width=0.95\linewidth]{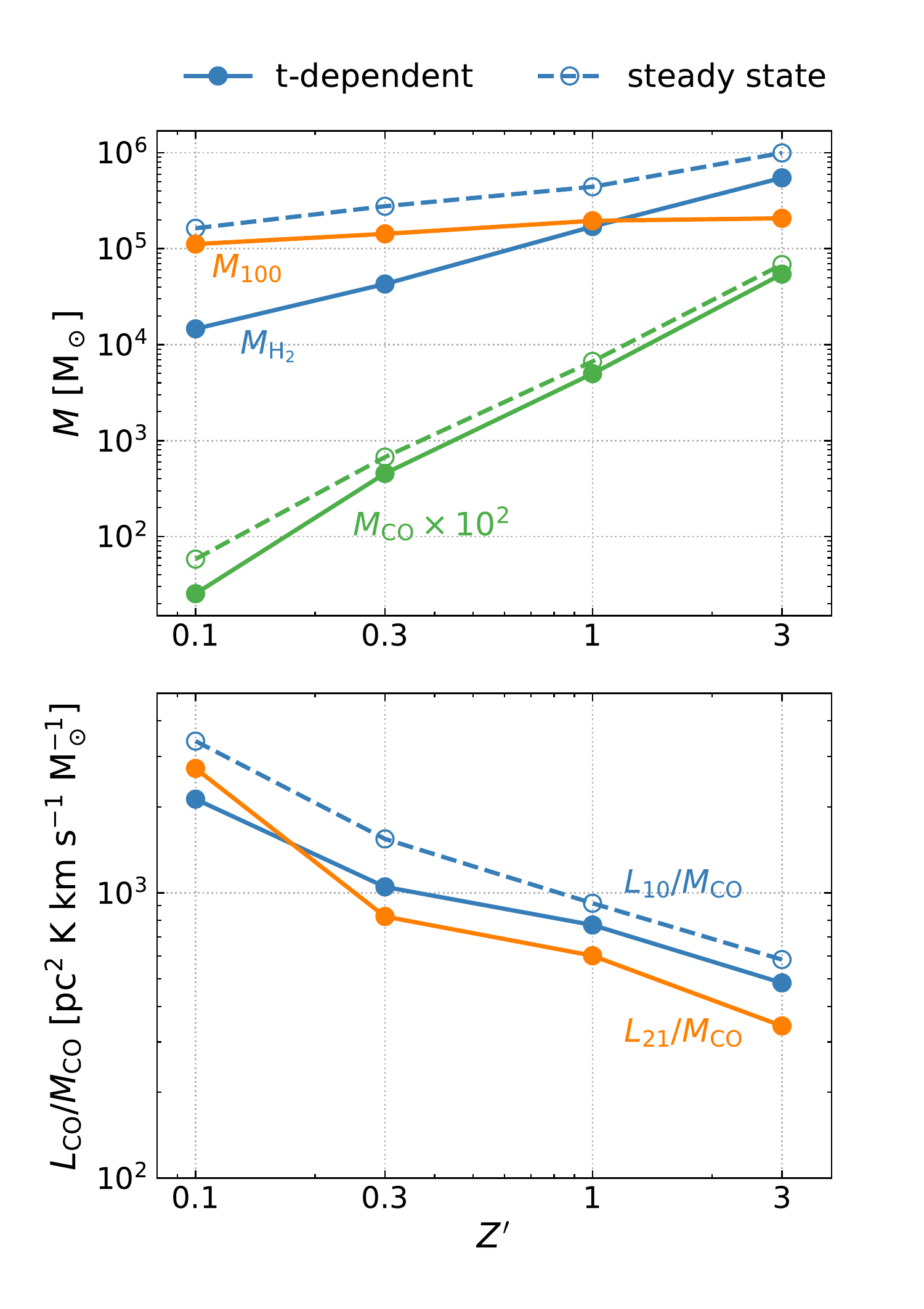}
	\caption{%
		\textit{Upper}: Total mass of H$_2$ (blue), CO (green, multiplied by 100), and dense gas ($n>100~{\rm cm^{-3}}$, orange) as functions of~$Z^\prime$. \textit{Lower}: Luminosity-to-mass ratio of \coone\ (blue) and \cotwo\ (orange) as functions of $Z^\prime$. The time-dependent and steady-state models are shown in solid and dashed lines, respectively.		
	}
	\label{fig:Z_vs_mass}
\end{figure}

As shown in \citetalias{HSvD21}, due to the long timescale of H$_2$ formation on dust grains, the dynamical effect significantly reduces the H$_2$ abundances leading to a lower total H$_2$ mass ($M_{\rm H_2}$) in the time-dependent model compared to its steady-state counterpart. In contrast, the effect on the CO abundance and thus the total CO mass ($M_{\rm CO}$) is much weaker. This is demonstrated in the upper panel of Fig.~\ref{fig:Z_vs_mass}. In addition, $M_{100}$ represents the total mass of the ``dense gas,'' which we define as $n > 100~{\rm cm^{-3}}$, and it is nearly independent of $Z^\prime$, consistent with the fact that the average star formation rate in \citetalias{HSvD21} is insensitive to~$Z^\prime$. Compared to $M_{\rm H_2}$, the star formation reservoir transitions from H$_2$-dominated at high $Z^\prime$ to \hi-dominated at low~$Z^\prime$.

While $M_{\rm H_2}$, $M_{\rm CO}$, and $M_{100}$ are purely chemical and hydrodynamical properties, the CO luminosity-to-mass ratio ($L_{\rm CO} / M_{\rm CO}$, lower panel of Fig.~\ref{fig:Z_vs_mass}) is a result of radiative transfer. The luminosities of \coone\ and \cotwo\ are defined as $L_{10} = \int W_{10}\,\mathrm{d}a$ and $L_{21} = \int W_{21}\,\mathrm{d}a$, respectively, where the integration is over the (1~kpc)$^2$ area. The luminosity-to-mass ratio is higher at low~$Z^\prime$ mainly because the lines become increasingly optically thin, as will be discussed in Section~\ref{sec:opt_depth}.

The time-averaged global conversion factor is $X_{\rm CO} = M_{\rm H_2} / (2 m_{\rm p} L_{10})$ where $m_{\rm p}$ is the proton mass.
We can now put these computed values in Fig.~\ref{fig:intro_global_Z_XCO} and compare them against observations.
The red filled circles and non-filled triangles represent our time-dependent model and steady-state model, respectively. The red non-filled stars show the ``dense gas'' conversion factor: $X_{100} \equiv N_{\rm H} (n > 100~{\rm cm^{-3}}) / W_{10} = 0.71 M_{100} / (2 m_{\rm p} L_{10})$, where the factor~$0.71$ is the hydrogen mass fraction.
Our fiducial time-dependent model 
can be fitted by (with a correlation coefficient of 0.99):
\begin{equation}\label{eq:fit1}
X_{\rm CO} = 3.17 \times 10^{20} Z^{\prime -0.71}\ {\rm cm^{-2}\ (K~km~s^{-1})^{-1}},
\end{equation}
which agrees well with the Milky Way value (Eq.~(\ref{eq:XCO_MW})) at  $Z^\prime = 1$ within 60\%.
The metallicity dependence is flatter compared to the steady-state counterpart where $X_{\rm CO} \propto Z^{\prime -1}$. This is because the dynamical effect mainly suppresses H$_2$ formation but not CO formation. The dense gas conversion factor scales even more steeply as $X_{100} \propto Z^{\prime -1.5}$. This demonstrates that the dense, star-forming gas reservoir becomes \hi-dominated at low~$Z^\prime$. The implication is that $X_{\rm CO}$ at low $Z^\prime$ can be overestimated both by the spectral synthesis method (by assuming steady-state chemistry) and by the inverse KS~method (by assuming the star-forming gas is fully molecular).

At super-solar metallicity ($Z^\prime = 3$),
our steady-state model agrees very well with our fiducial time-dependent model.
Steady-state chemistry is a good approximation as the H$_2$ formation time is short.
However,
the dense gas conversion factor is significantly lower than $X_{\rm CO}$,
which reflects the fact that a large fraction of diffuse (and presumably cold) gas is molecular.

\subsection{Visual Impression}

\begin{figure*}
	\centering
	\includegraphics[width=0.99\linewidth]{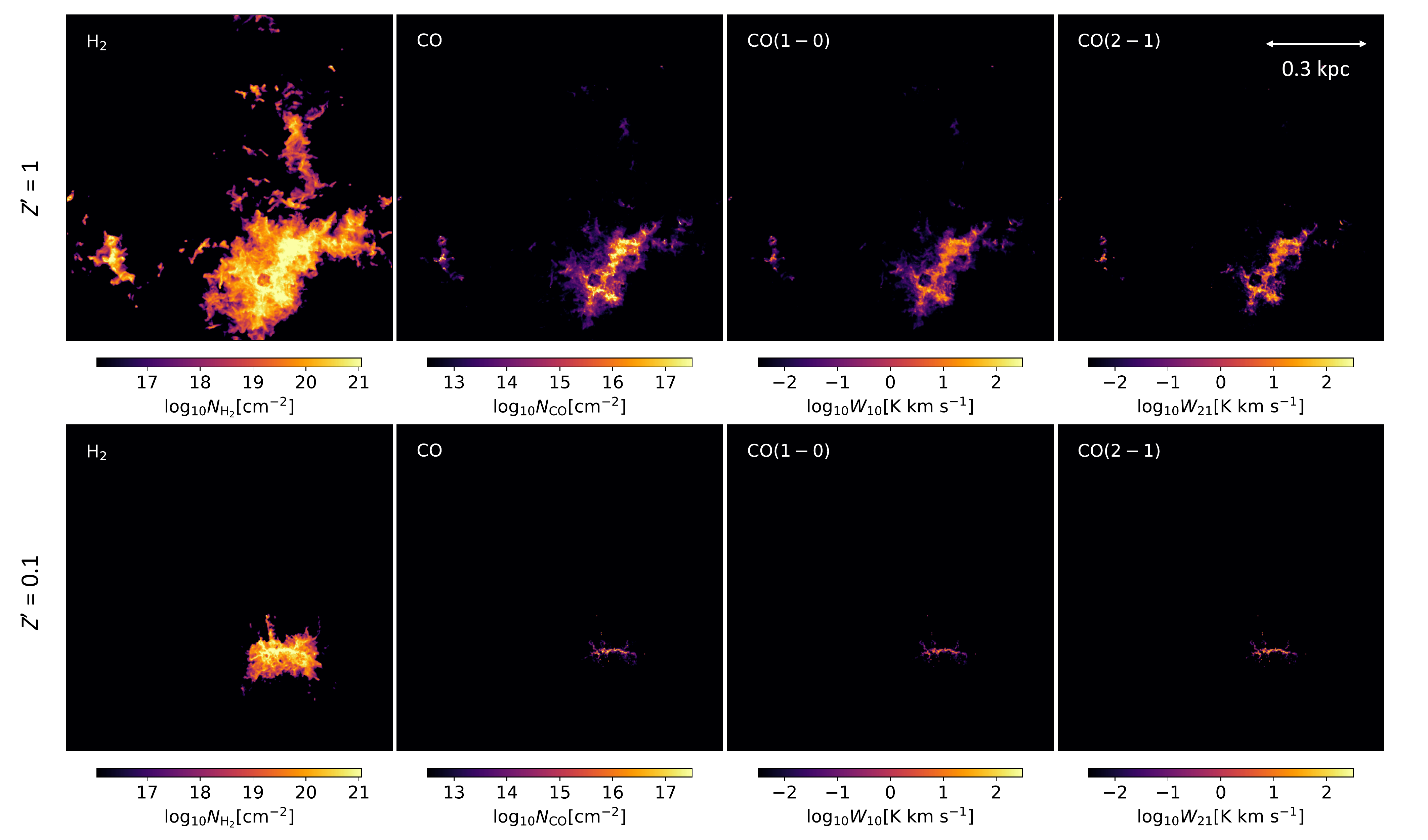}
	\caption{%
		Face-on maps of H$_2$ column density, CO column density, \coone\ line intensity, and \cotwo\ line intensity from left to right in the solar metallicity ($Z^\prime = 1$) run at $t = 420$~Myr (upper row) and the $Z^\prime = 0.1$ run at $t = 130$~Myr (lower row). The entire simulation domain of 1~kpc$^2$ is shown.
		H$_2$ is spatially more extended than CO. Both H$_2$ and CO are more compact at low metallicity.
	}
	\label{fig:co_10_21_maps}
\end{figure*}

Fig.~\ref{fig:co_10_21_maps} shows face-on maps%
\footnote{%
    We note that \citetalias{HSvD21} distinguishes between the line-of-sight integrated column density ($N^{\rm obs}$) and the column density available for radiation shielding ($N^{\rm eff}$). In this work, we refer by column density exclusively to $N^{\rm obs}$.}
of $N_{\rm H_2}$, $N_{\rm CO}$, $W_{10}$, and $W_{21}$ from left to right
with a beam size of $l_{\rm b} = 2$~pc
in the $Z^\prime = 1$ run at $t = 420$~Myr (upper row) and the $Z^\prime = 0.1$ run at $t = 130$~Myr (lower row). The entire simulation domain of 1~kpc$^2$ is shown.
Qualitatively, CO only exists in the inner part of the H$_2$~clouds where the gas is dense and well shielded. 
Both H$_2$ and CO are more compact in the $Z^\prime = 0.1$ case.
The distribution of $W_{10}$ is very similar to that of $N_{\rm CO}$ except at the densest cores where $W_{10}$ is slightly reduced compared to $N_{\rm CO}$. This is because \coone\ transitions from the optically thin regime where $W_{10} \propto N_{\rm CO}$ to the optically thick regime where $W_{10}$ saturates. $W_{21}$ follows a similar distribution to $W_{10}$ but is less spatially extended, as the $J = 2$ level is not sufficiently excited in the diffuse gas.

However, for extragalactic observations, a pc-scale resolution is often inaccessible.
Therefore, we construct $N_{\rm H_2}$, $N_{\rm CO}$, $W_{10}$, and $W_{21}$ at coarser beam sizes from our results for $l_{\rm b} = 2$~pc. The beam size is systematically increased by factors of two up to $l_{\rm b} = 1$~kpc, which includes the entire simulation domain. For example, $N_{\rm H_2}$ and $W_{10}$ at $l_{\rm b} = 4$~pc are constructed from the results at $l_{\rm b} = 2$~pc:
\begin{align}
    N_{\rm H_2}\,(4~{\rm pc}) &= \frac{\int N_{\rm H_2}\,\mathrm{d}a}{\int \mathrm{d}a} = \frac{1}{2^2} \sum_{i=1}^{4} N_{{\rm H_2},i}\,(2~{\rm pc})~,\\
    W_{10}\,(4~{\rm pc}) &= \frac{\int W_{10}\,\mathrm{d}a}{\int \mathrm{d}a} = \frac{1}{2^2} \sum_{i=1}^{4} W_{10,i}\,(2~{\rm pc})~,
\end{align}
where the integration is over the area of a $4$~pc beam and the summation is over the 2-by-2 sub-beams of $l_{\rm b} = 2$~pc which constitute a $4$~pc beam. Likewise, a $8$~pc beam is constructed from the 4-by-4 sub-beams of $2$~pc (or, equivalently, from the 2-by-2 sub-beams of $4$~pc), and so forth.

\begin{figure*}
	\centering
	\includegraphics[width=0.99\linewidth]{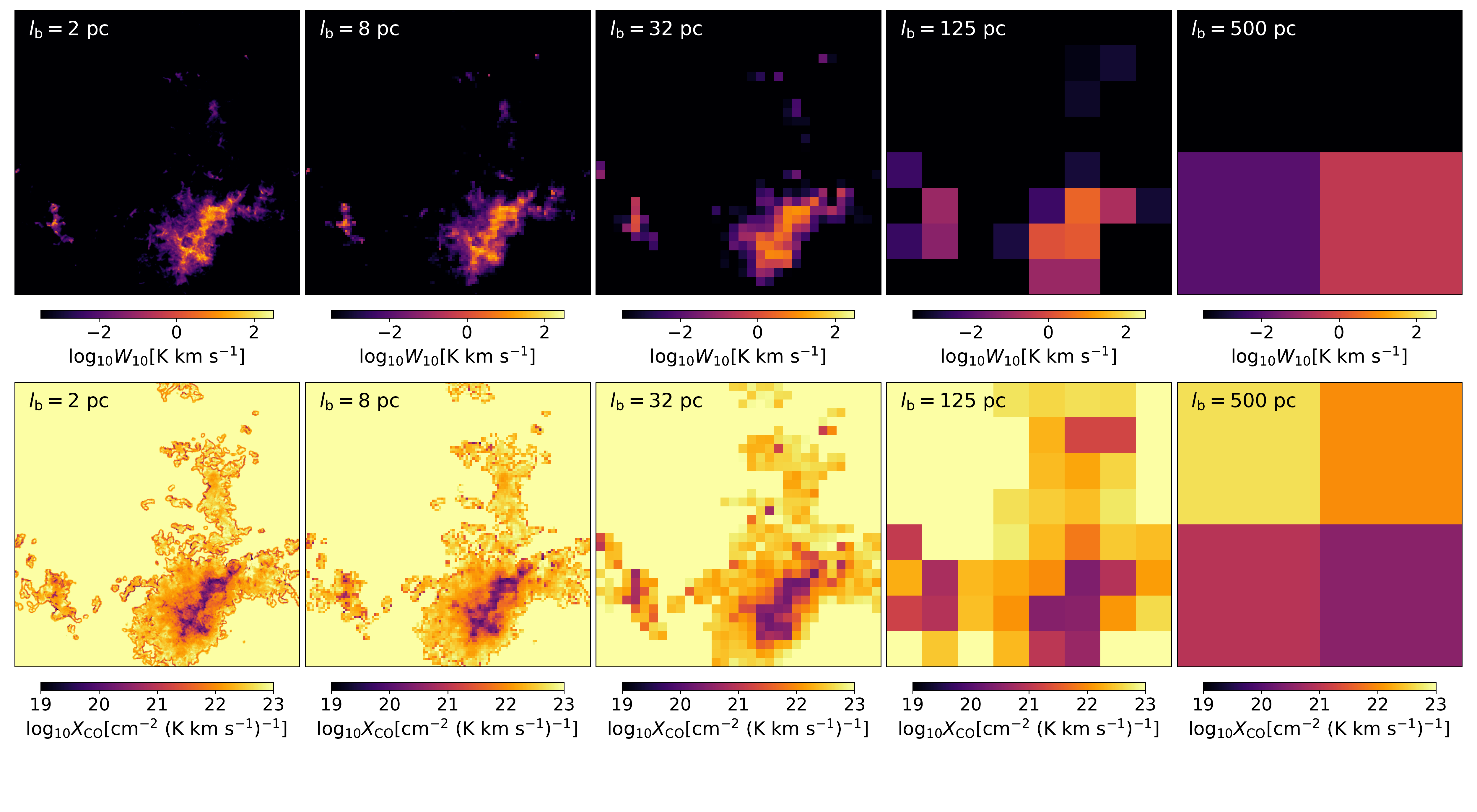}
	\caption{
		Face-on maps of $W_{10}$ (upper row) and $X_{\rm CO}$ (lower row)
		with beam size $l_{\rm b} = 2, 8, 32, 125, \text{and } 500$~pc from left to right. Note that the Milky Way $X_{\rm CO}$ factor (Eq.~\eqref{eq:XCO_MW}) is represented by the dark purple color.}
	\label{fig:XCO_maps_pix}
\end{figure*}

To illustrate the effect of spatial averaging/\linebreak[0]{}smoothing, Fig.~\ref{fig:XCO_maps_pix} shows the maps of $W_{10}$ (upper row) and $X_{\rm CO}$ (lower row) for $l_{\rm b} = 2, 8, 32, 125, \text{and } 500$~pc from left to right. On $2$~pc scale, $X_{\rm CO}$ varies by orders of magnitude: it is low in the dense, well-shielded gas and high in the diffuse gas. We stress that our radiative transfer calculations are done with a much smaller minimal cell size $\sim 0.2$~pc with the sub-pixeling technique (see Section~\ref{sec:AMR}).
On large scales, as long as the beam contains dense gas (which only occupies a small area), the beam-averaged $X_{\rm CO}$ is driven towards low values (in purple). This will be discussed more quantitatively in Section~\ref{sec:obs_impl} after we provide more insight and background information on the simulations at the molecular level in Section~\ref{sec:mol_physics}. 

\subsection{Chemical Properties}

In this section, we briefly describe the distributions of H$_2$ and CO in the simulations. Although these chemical properties have been shown in \citetalias{HSvD21}, they are the basis of our radiative transfer and provide context for the results in Fig.~\ref{fig:XCO_maps_pix},
so they are worth repeating here.

\begin{figure*}
	\centering
	\includegraphics[width=0.99\linewidth]{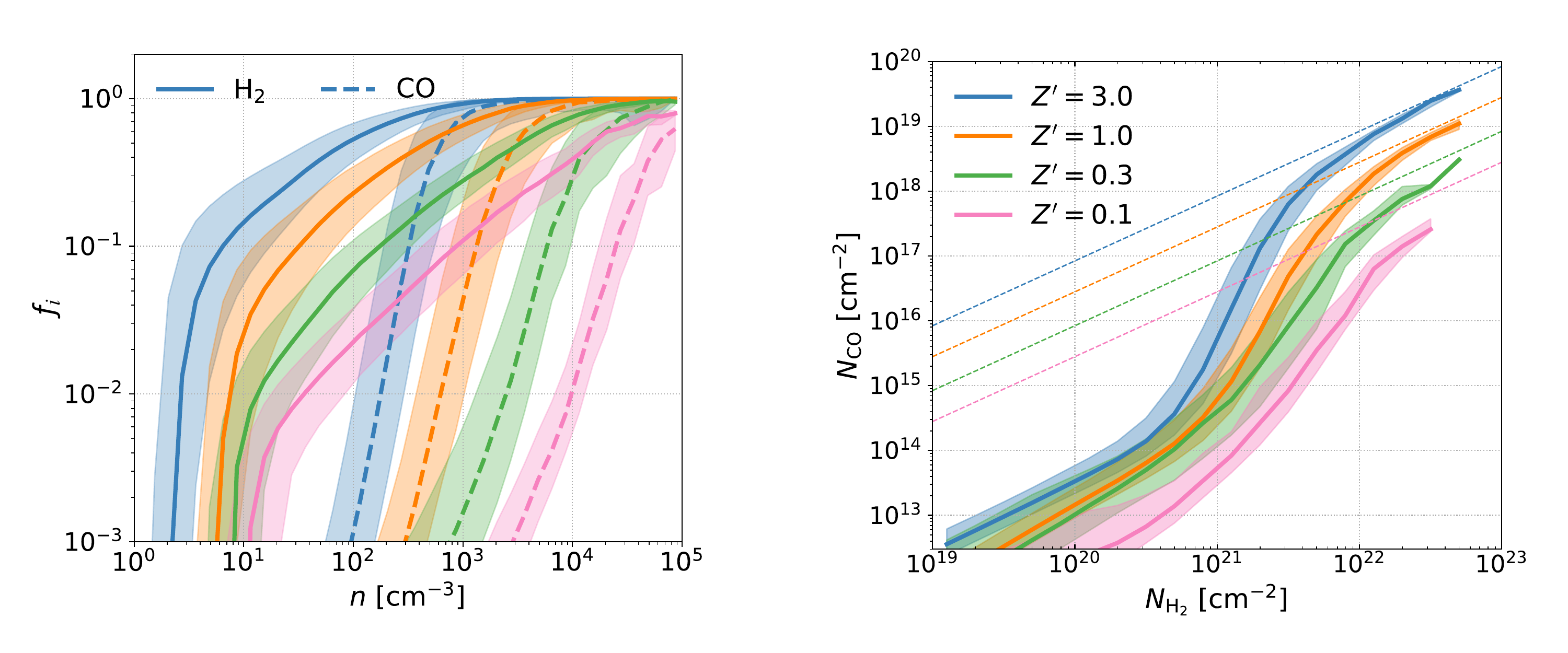}
	\caption{%
		\textit{Left}: Normalized fractional abundance of H$_2$ (solid lines) and CO (dashed lines) as functions of hydrogen number density~$n$ for normalized metallicity $Z^{\prime} = 3, 1, 0.3, \text{and } 0.1$. \textit{Right}: H$_2$~column density vs.\ CO column density for $Z^{\prime} = 3, 1, 0.3, \text{and } 0.1$. 
		The thin dashed lines indicate the maximum ratio $N_{\rm CO}/N_{\rm H_2} = 2 x_{\rm C,0}$ where carbon is completely in form of CO.
		In both panels, the lines represent the median values while the shaded areas enclose the 25th and 75th percentiles.
	}
	\label{fig:summary_chem}
\end{figure*}

We define the number abundance of a chemical species $i$ as $x_i\equiv n_i / n$, where $n$ is the hydrogen number density and $n_i$ is the number density of species $i$. The normalized fractional abundance of H$_2$ is then defined as $f_{\rm H_2} \equiv 2x_{\rm H_2}$ such that $f_{\rm H_2} = 1$ when all hydrogen is in the form of H$_2$. Similarly, the normalized fractional abundance of CO is defined as $f_{\rm CO} \equiv x_{\rm CO} / x_{\rm C,0}$, where $x_{\rm C,0} = 1.4\times 10^{-4} Z^\prime$ is the total carbon abundance. The left panel of Fig.~\ref{fig:summary_chem}, shows $f_{\rm H_2}$ (solid lines) and $f_{\rm CO}$ (dashed lines) as functions $n$. At a given $Z^\prime$, CO forms at higher densities than where H$_2$ forms, consistent with the visual impression in Fig.~\ref{fig:co_10_21_maps}. Both H$_2$ and CO exist at increasingly higher densities as $Z^\prime$ decreases due to the deficit of dust.

The right panel of Fig.~\ref{fig:summary_chem} shows $N_{\rm CO}$ as a function of $N_{\rm H_2}$. The $N_{\rm H_2}{-}N_{\rm CO}$ relation is characterized by three regimes: (i)~The CO-poor regime at low $N_{\rm H_2}$ where the CO-to-H$_2$ ratio $\gamma \equiv N_{\rm CO}/N_{\rm H_2}$ is low; (ii)~the transition regime at intermediate $N_{\rm H_2}$ where $\gamma$ increases sharply with $N_{\rm H_2}$ which reflects the formation of CO; and (iii)~the CO-rich regime at high $N_{\rm H_2}$ where $\gamma$ approaches its upper limit (indicated by the thin dashed lines).


\begin{figure}
	\centering
	\includegraphics[trim=0 0cm 0 1cm,clip,width=0.95\linewidth]{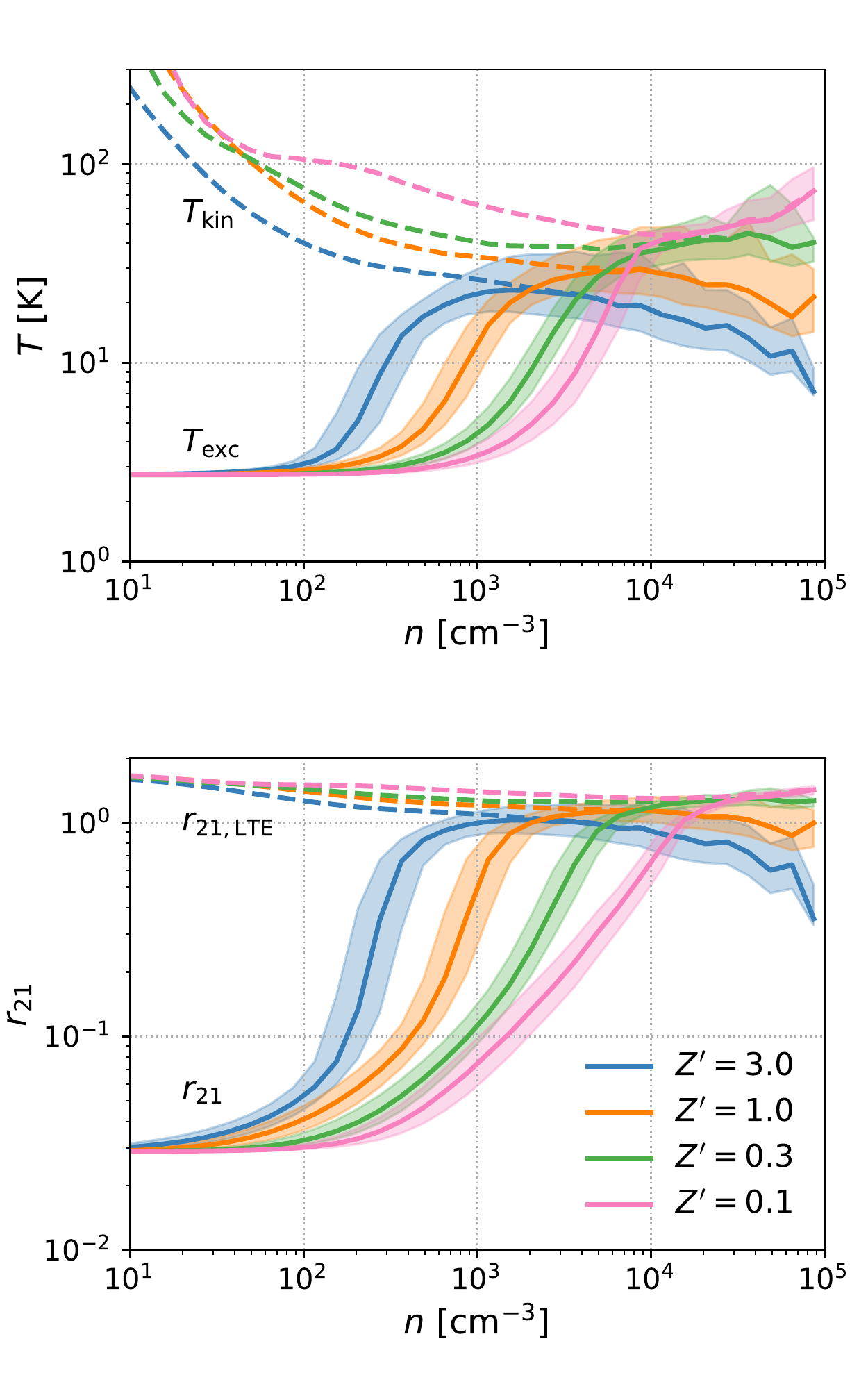}
	\caption{%
		\textit{Upper panel}: Excitation temperature of \coone\ ($T_{\rm exc}$, solid lines) and kinetic temperature ($T_{\rm kin}$, dashed lines) as a function of~$n$. \textit{Lower panel}: CO population ratio of $J = 2$ to $J = 1$ ($r_{21}$, solid lines) as a function of~$n$. The dashed lines show the CO population ratio assuming local thermodynamic equilibrium ($r_{\rm 21,LTE}$). The lines represent the median values while the shaded areas enclose the 25th and 75th percentiles (not shown for $T_{\rm kin}$ and $r_{\rm 21,LTE}$ for the sake of clarity).
		The lines thermalize at higher densities at low $Z^\prime$ as 
		the collision partner H$_2$ forms at densities higher than the critical densities
		and the lower CO abundance leads to less efficient radiation trapping.
	}
	\label{fig:nH_vs_x_T_R21}
\end{figure}

\section{Molecular physics}\label{sec:mol_physics}

In this section,
we present a detailed analysis of the ``microphysics'' in our simulations and discuss how the lines are excited and propagated in the ISM, which forms the backbone of the observables.
The readers interested only in our predicted $X_{\rm CO}$ and $R_{21}$ may want to skip this section and go directly to Section~\ref{sec:obs_impl}.

\subsection{CO Excitation}\label{sec:excitation}

In the upper panel of Fig.~\ref{fig:nH_vs_x_T_R21}, we show the excitation temperature of \coone\ ($T_{\rm exc}$, solid lines) and kinetic temperature ($T_{\rm kin}$, dashed lines) as a function of~$n$.
In low-density gas where collisions are unimportant, $T_{\rm exc}$ is set by the background radiation $T_{\rm bg} = 2.73$~K due to the cosmic microwave background (CMB). As $n$ increases, $T_{\rm exc}$ gradually increases and approaches $T_{\rm kin}$, and the line is ``thermalized'' when $T_{\rm exc} = T_{\rm kin}$. At the highest densities, $T_{\rm kin}$ (and therefore $T_{\rm exc}$) is higher at low $Z^\prime$ due to heating from H$_2$ formation and UV pumping \citep{2019ApJ...881..160B}. Interestingly, \coone\ is thermalized at higher densities as $Z^\prime$ decreases. This is caused by a combination of two effects at low $Z^\prime$. Firstly, the collision partner H$_2$ forms at densities higher than the critical density of \coone. Therefore, it takes a higher total hydrogen number density to thermalize the line.
Secondly, higher CO abundance results in more efficient radiation trapping (more optically thick) and thus a lower effective critical density.

The CO population ratio~($r_{21}$) of level $J = 2$ to level $J = 1$ follows a similar trend. In the lower panels of Fig.~\ref{fig:nH_vs_x_T_R21}, we show $r_{21}$  as a function of~$n$ (left) and $n_{\rm H_2}$ (right) in solid lines. Analytically, the population ratio of level~$J$ to level $J^\prime = J-1$ can be expressed as a function of $T_{\rm exc}$:
\begin{eqnarray}\label{eq:rJJanalytic}
	r_{JJ^\prime}(T_{\rm exc})
	&=& \frac{ (2J + 1) e^{-B_0 J(J+1) / T_{\rm exc}} }{ (2J^\prime + 1) e^{-B_0 J^\prime(J^\prime+1) / T_{\rm exc}} } \nonumber\\
	&=& \frac{g_J}{g_{J^\prime}} e^{- T_{JJ^\prime} / T_{\rm exc}}~, 
\end{eqnarray}
where $g_J$ and $g_{J^\prime}$ are the degeneracy in level $J$ and $J^\prime$, respectively, $B_0 = 2.77$~K, $T_{JJ^\prime} = 2 J B_0 = E_{JJ^\prime} / k_\mathrm{B}$, and $E_{JJ^\prime}$ is the energy difference between levels $J$ and $J^{\prime}$. For $J=2$ and $J^{\prime}=1$, it follows
\begin{equation}\label{eq:r21analytic}
    r_{21}(T_{\rm exc}) = 
\frac{5}{3}e^{-11.1~{\rm K} / T_{\rm exc}}~.
\end{equation}
Note that $T_{\rm exc}$ here is specifically for \cotwo\ which can be different from the excitation temperature of \coone\ unless the gas is in LTE. At low densities where $T_{\rm exc}=2.73$~K, our results agree well with the analytic expectation $r_{21}(2.73~{\rm K}) = 0.029$. As $n$  increases, $r_{21}$ gradually increases and approaches the LTE line ratio $r_{\rm 21,LTE}$ (dashed lines), which we obtain using Eq.~\eqref{eq:r21analytic} assuming $T_{\rm exc} = T_{\rm kin}$. At $Z^\prime \geq 1$, the decline of $r_{21}$ at high~$n$ is due to the decline of $T_{\rm kin}$ instead of subthermal excitation. The maximum $r_{21}$ is around unity at all metallicities and is slightly higher at low $Z^\prime$ due to higher $T_{\rm kin}$, though the effect is very mild.
Indeed, Eq.~\eqref{eq:r21analytic} suggests that $r_{21}$ is not very sensitive to $T_{\rm exc}$ when $T_{\rm exc} \gg 11.1$~K and it asymptotes to an upper limit of $5/3 \approx 1.66$ as $T_{\rm exc} \to \infty$.

\subsection{Optically Thin Limit}

\begin{figure*}
	\centering
	\includegraphics[width=1.\linewidth]{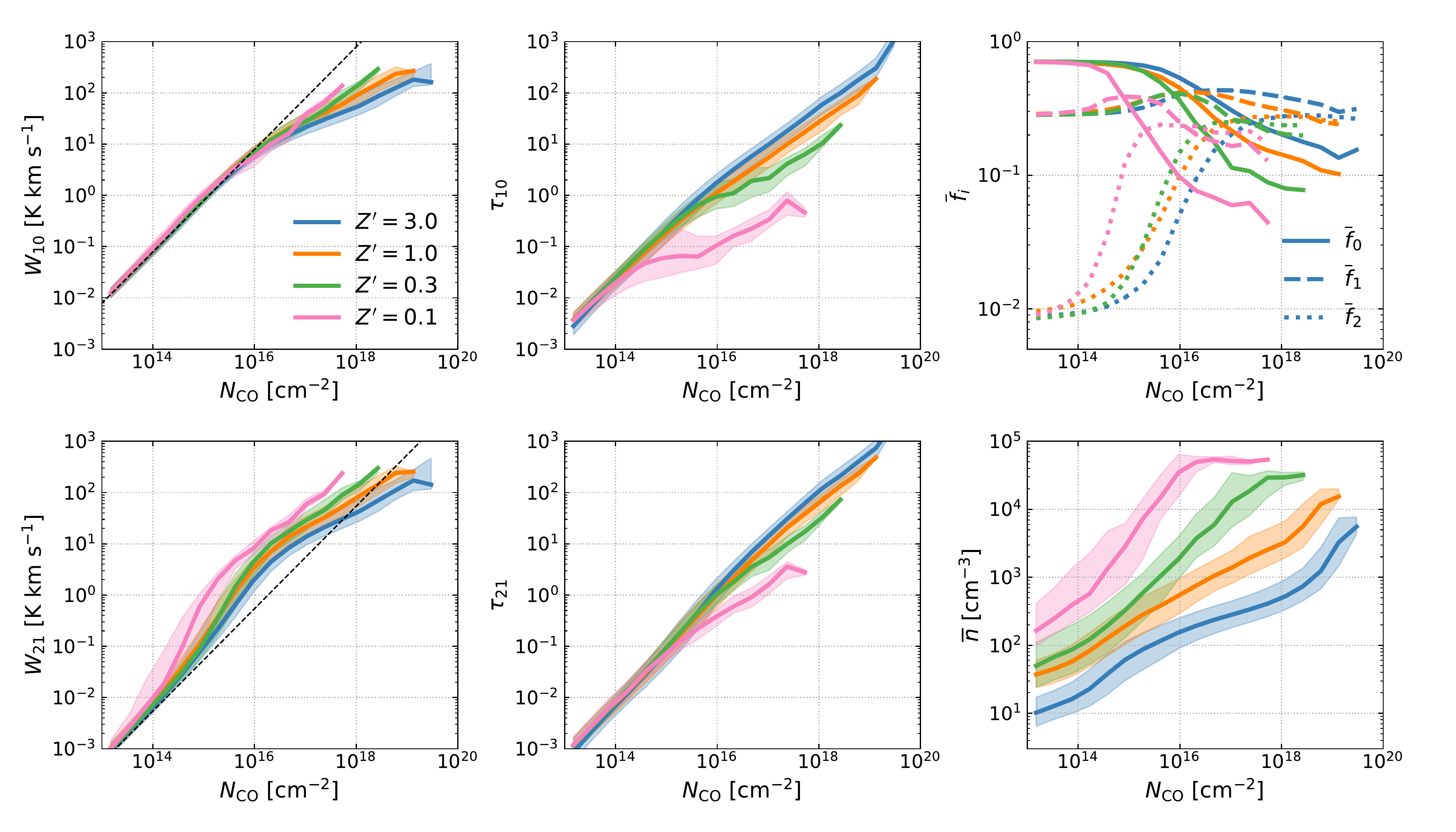}
	\caption{%
		\textit{Left}: Line intensity of \coone\ (upper) and \cotwo\ (lower) as a function of $N_{\rm CO}$. The dashed lines in the upper and lower panels indicate $W_{\rm 10,thin} / N_{\rm CO,16} = 7.82\ {\rm K~km~s^{-1}} $ and $W_{\rm 21,thin} /  N_{\rm CO,16} = 0.54\ {\rm K~km~s^{-1}} $, respectively. \textit{Middle}: Optical depth of \coone\ (upper) and \cotwo\ (lower) as a function of $N_{\rm CO}$. \textit{Upper right}: The $n_{\rm CO}$-weighted average fraction of CO in the rotational level $J = 0$ (solid), 1~(dashed), and 2 (dotted) as functions of $N_{\rm CO}$. \textit{Lower right}: The $n_{\rm CO}$-weighted average hydrogen number density as a function of $N_{\rm CO}$. The lines represent the median values while the shaded areas enclose the 25th and 75th percentiles in all panels (not shown for $\overline{f}_J$  for the sake of clarity).		
	}
	\label{fig:nco_vs_tau_wco}
\end{figure*}

At low $N_{\rm CO}$, the lines are in the optically thin limit and self-absorption can be neglected. The emissivity of the CO($J{-}J^\prime$) transition is
\begin{equation}
    j_\nu = \frac{E_{JJ^\prime}}{4\pi}n_J A_{JJ^\prime} \phi(\nu)~,
\end{equation}
where $n_J$ is the number density of CO in level~$J$ and $A_{JJ^\prime}$ is the Einstein coefficient for spontaneous decay.
As there is no absorption, the specific intensity is simply $I_\nu = \int j_\nu\,\mathrm{d}z$ and thus the line intensity can be written as
\begin{eqnarray}
    W_{JJ^\prime,{\rm thin}} &=& \frac{\lambda_{JJ^\prime}^2}{2 k_\mathrm{B}} \iint j_\nu\,\mathrm{d}z\,\mathrm{d}v \nonumber\\
    &=& \frac{\lambda_{JJ^\prime}^3}{8\pi}T_{JJ^\prime}A_{JJ^\prime} \int n_{\rm CO} f_J\,\mathrm{d}z~,
\end{eqnarray}
where 
$\lambda_{JJ^\prime} = h c / E_{JJ^\prime}$ is the line-center wavelength,  $h$~is the Planck constant, and $f_J\equiv n_J / n_{\rm CO}$ is the fraction of CO in level~$J$. If we define the $n_{\rm CO}$-weighted line-of-sight (l.o.s.) average $f_J$ as
\begin{equation}
    \overline{f}_J  \equiv \frac{\int n_{\rm CO} f_J\,\mathrm{d}z}{ \int n_{\rm CO}\,\mathrm{d}z }
    = \frac{N_{{\rm CO},J} }{N_{\rm CO}}~,
\end{equation}
where $N_{\rm CO}$ is the CO column density and $N_{{\rm CO},J}$ is the column density of CO in level~$J$, we obtain
\begin{equation}\label{eq:Wthin}
    W_{JJ^\prime,{\rm thin}} = \frac{\lambda_{JJ^\prime}^3}{8\pi}T_{JJ^\prime}A_{JJ^\prime} N_{\rm CO}  \overline{f}_J~.
\end{equation}
Namely, $W_{JJ^\prime,{\rm thin}}$ scales linearly with $N_{\rm CO}$, with a secondary dependency on~$\overline{f}_J$. Note that Eq.~\eqref{eq:Wthin} is independent of the line width~$b$ as expected for optically thin conditions.


Analytically, $f_J$~can be expressed as a function of the excitation temperature $T_{\rm exc}$
\begin{eqnarray}\label{eq:fJanalytic}
    f_J(T_{\rm exc}) 
    &=& \frac{ (2J + 1) e^{-B_0 J(J+1) / T_{\rm exc}} }{ \sum_J (2J + 1) e^{-B_0 J(J+1) / T_{\rm exc}}  } \nonumber\\
    &\approx& \frac{ (2J + 1) e^{-B_0 J(J+1) / T_{\rm exc}} }{\sqrt{1 + (T_{\rm exc} / B_0)^2}}~,
\end{eqnarray}
where the approximation of the partition function in the denominator is accurate to within $\pm 6\%$ for all $T_{\rm exc}$ (\citealp{2011piim.book.....D}, ch.~19).
Adopting
$A_{10} = 7.2\times 10^{-8}~{\rm s^{-1}}$,
$A_{21} = 6.91\times 10^{-7}~{\rm s^{-1}}$,
$T_{10} = 5.53$~K,
$T_{21} = 11.1$~K,
$\lambda_{10} = 0.26$~cm,
and $\lambda_{21} = 0.13$~cm,
the line intensities in the optically thin regime become
\begin{align}
    W_{\rm 10,thin} &= 7.82~{\rm K~km~s^{-1}} N_{\rm CO,16} \frac{\overline{f}_1 }{f_1(2.73)}~,\label{eq:W10_thin}\\
    W_{\rm 21,thin} &= 0.54~{\rm K~km~s^{-1}} N_{\rm CO,16} \frac{\overline{f}_2 }{f_2(2.73)}~,\label{eq:W21_thin}
\end{align}
where $N_{\rm CO,16} \equiv N_{\rm CO}/ (10^{16}~{\rm cm^{-2}})$.

In the left panels in Fig.~\ref{fig:nco_vs_tau_wco}, we show $W_{10}$ (upper) and $W_{21}$ (lower) as functions of $N_{\rm CO}$ at different~$Z^\prime$. At the lowest $N_{\rm CO}$ where collisional excitation is inefficient, $T_{\rm exc} = 2.73$~K and thus $W_{10}$ and $W_{21}$ follow the dashed lines which indicate $W_{\rm 10}  = 7.82 N_{\rm CO,16}\ {\rm K~km~s^{-1}}$ and $W_{\rm 21}  = 0.54  N_{\rm CO,16}\ {\rm K~km~s^{-1}}$ in the upper and lower panels, respectively. For $N_{\rm CO} < 10^{16}~{\rm cm^{-3}}$, $W_{10}$ scales linearly with $N_{\rm CO}$
as the secondary factor $\overline{f}_1$ does not vary much with~$N_{\rm CO}$. This is shown in the upper right panel of Fig.~\ref{fig:nco_vs_tau_wco}. Since the $J = 1$ level is only $5.53$~K above the ground state, it is already significantly excited by the CMB radiation alone ($f_1\,(2.73~{\rm K}) = 0.28$). In contrast, $W_{21}$ scales super-linearly with $N_{\rm CO}$ in the optically thin regime where $\overline{f}_2$ provides a secondary contribution.
Indeed, $\overline{f}_2$ increases significantly with $N_{\rm CO}$, from $f_2\,(2.73~{\rm K}) = 8.08\times 10^{-3}$ to a maximum value of ${\sim}0.3$.

In addition, the excitation of both $J = 1$ and $J = 2$ occurs at lower $N_{\rm CO}$ as $Z^\prime$ decreases. Assuming a positive correlation between~$n$ and~$N_{\rm CO}$, this seems to be in conflict with Fig.~\ref{fig:nH_vs_x_T_R21} where thermalization occurs at higher densities at low $Z^\prime$. However, since CO only exists in the small and dense cores at low $Z^\prime$, the corresponding gas density is higher at a given $N_{\rm CO}$. This is demonstrated in the lower right panel of Fig.~\ref{fig:nco_vs_tau_wco}, where we show the $n_{\rm CO}$-weighted l.o.s.\ average hydrogen number density,
\begin{equation}
    \overline{n} \equiv \frac{\int n_{\rm CO} n\,\mathrm{d}z}{\int n_{\rm CO}\,\mathrm{d}z}~,
\end{equation}
as a function of~$N_{\rm CO}$. Indeed, $\overline{n}$ correlates positively with $N_{\rm CO}$ at all~$Z^\prime$. However, at a given $N_{\rm CO}$, $\overline{n}$ increases inversely with~$Z^\prime$. The net effect is that as $Z^\prime$ decreases the excitation of both $J = 1$ and $J = 2$ occurs at a lower~$N_{\rm CO}$.

\subsection{Optical Depth Effect}\label{sec:opt_depth}


The optical depth at the line center of CO($J{-}J^\prime$) can be expressed as
\begin{equation}\label{eq:tauJJ}
    \tau_{JJ^\prime} = \frac{\lambda_{JJ^\prime}^3 A_{JJ^\prime} g_J}{8\pi^{3/2} g_{J^\prime}} \int \frac{n_{J^\prime}}{b} \left(1 - \frac{n_J g_{J^\prime}}{n_{J^\prime} g_J} \right) \mathrm{d}z~.
\end{equation}
Adopting $g_0 = 1$, $g_1 = 3$, and $g_2 = 5$, we obtain
\begin{align}
    \tau_{10} 
    &= 8.53 N_{\rm CO,16} \overline{ \frac{f_0}{b_5}(1 - \frac{ r_{10}}{3}) } \label{eq:tau10}\\
\intertext{and}
    \tau_{21} 
    &= 5.68 N_{\rm CO,16} \overline{ \frac{f_1}{b_5}(1 - \frac{ 3 r_{21}}{5}) }~,\label{eq:tau21}
\end{align}
where $b_5 \equiv b/(10^5~{\rm cm~s^{-1}})$. We numerically calculate the $n_{\rm CO}$-weighted l.o.s.\ average terms in Eqs.~\eqref{eq:tau10} and~\eqref{eq:tau21} and show them as functions of $N_{\rm CO}$ in the middle panels of Fig.~\ref{fig:nco_vs_tau_wco}.

Adopting characteristic values $b_5 = 2$ and $T_{\rm exc} = 10$~K, we expect $\tau_{10} = 1$ at $N_{\rm CO} \sim 2.1\times 10^{16}~{\rm cm^{-2}}$ and $\tau_{21} = 1$ at $N_{\rm CO} \sim 1.2\times 10^{16}~{\rm cm^{-2}}$, respectively. This analytic estimate agrees well with our results (except for the $Z^\prime = 0.1$ case where the $T_{\rm exc}$ is much higher). At higher $N_{\rm CO}$, the lines become optically thick and thus both $W_{10}$ and $W_{21}$ start to flatten and scale sub-linearly with $N_{\rm CO}$.

At a given $N_{\rm CO}$, $\tau_{10}$ decreases with $Z^\prime$. This is caused by the corresponding higher density~($\overline{n}$) that more efficiently excites the upper level $J=1$, which means (i)~higher $r_{10}$ and thus more efficient stimulated emission and (ii)~lower $f_0$ as the ground level $J=0$ is de-populated. In other words, \coone\ becomes increasingly optically thin as $Z^\prime$ decreases. In particular, at $Z^\prime = 0.1$, $\tau_{10}$ never exceeds unity at all $N_{\rm CO}$.
On the other hand, $\tau_{21}$ shows a similar dependency on $Z^\prime$, but the effect is less significant than for $\tau_{10}$, as $f_1$ does not vary with $N_{\rm CO}$ as much as $f_0$. Therefore, at $Z^\prime = 0.1$, $\tau_{21}$ does eventually become optically thick, reaching a maximum of three at the highest $N_{\rm CO}$.

\begin{figure*}
	\centering
	\includegraphics[width=0.99\linewidth]{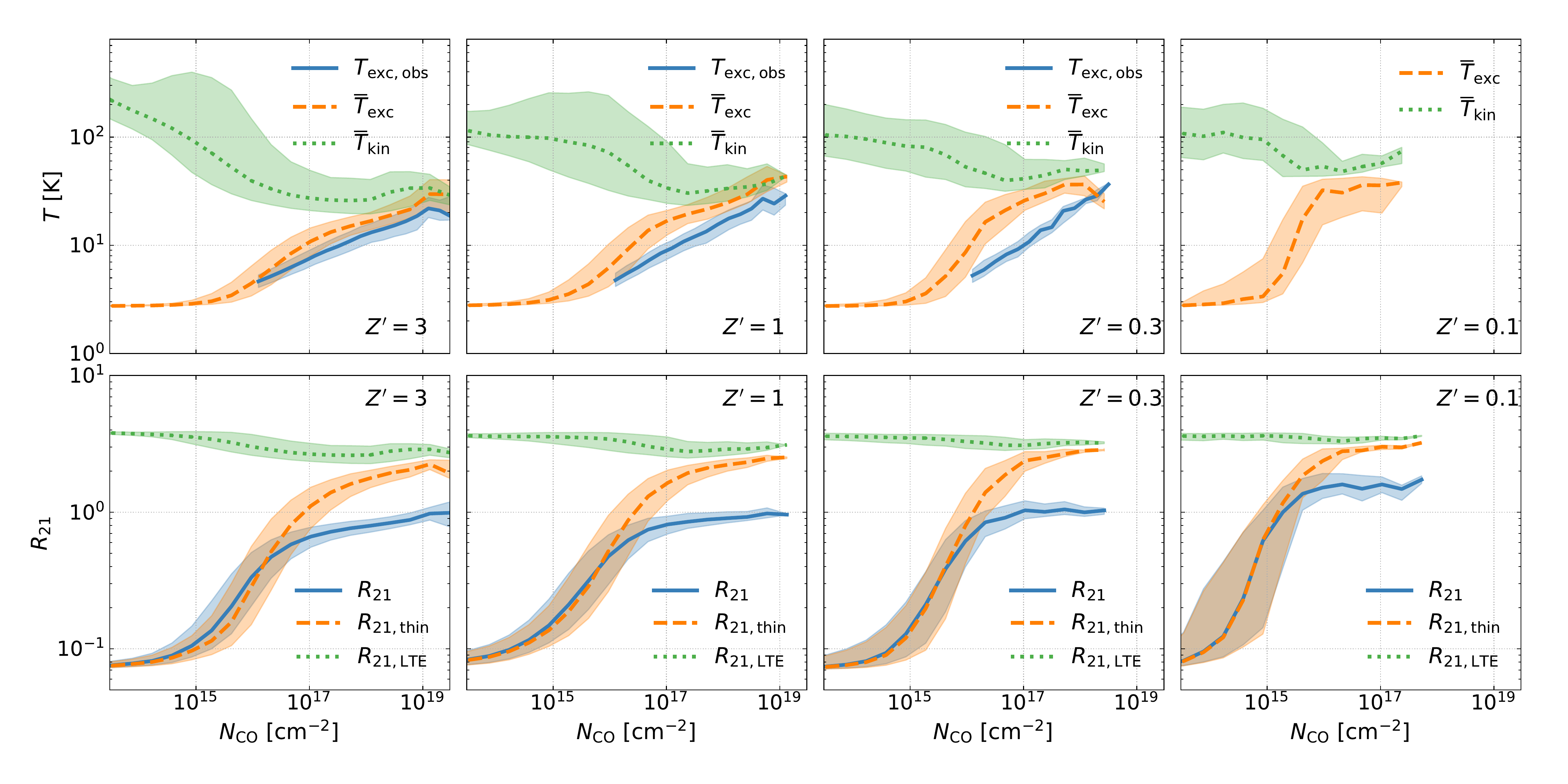}
	\caption{%
		\textit{Upper panels}: Cloud surface excitation temperature as would be observed (blue solid, see Eq.~\eqref{eq:Texc_obs}), CO density-weighted average excitation temperature (orange dashed, see Eq.~\eqref{eq:Texc_los_ave}), and CO density-weighted average kinetic temperature (green dotted, see Eq.~\eqref{eq:Tkin_los_ave}) as functions of $N_{\rm CO}$ for $Z^{\prime} = 3, 1, 0.3, \text{and } 0.1$ from left to right. \textit{Lower panels}: Observed line ratio $R_{21}\equiv W_{21}/W_{10}$ (blue solid), ``optically thin'' line ratio (orange dashed, see Eq.~\eqref{eq:Rthin}), and ``optically thin'' line ratio assuming LTE (green dotted, see Eq.~\eqref{eq:RthinLTE}) as functions of $N_{\rm CO}$ for $Z^{\prime} = 3, 1, 0.3, \text{and } 0.1$ from left to right. The lines represent the median values while the shaded areas enclose the 25th and 75th percentiles in all panels.
		The observed excitation temperature and line ratio are slightly lower than the l.o.s. averages as only the cloud surfaces can be seen for optically thick lines.
	}
	\label{fig:NCO_vs_temp_Z}
\end{figure*}

Observationally, the excitation temperature of \coone\ can be estimated from the peak of the observed radiation temperature $T_{\rm R,max}$ \citep[e.g.,][]{2008ApJ...679..481P}. This is based on the radiative transfer equation in a uniform medium:
\begin{equation}
	T_\mathrm{R} = T_{10} \Big( \frac{1}{e^{T_{10}/T_{\rm exc}} - 1} - \frac{1}{e^{T_{10}/T_{\rm bg}} - 1}\Big)(1 - e^{-\tau})~.
\end{equation}
If the line is sufficiently optically thick ($\tau \gg 1$), the excitation temperature can be expressed as
\begin{equation}\label{eq:Texc_obs}
	T_{\rm exc,s} = \frac{T_{10}}{ \ln(1 + T_{10} / \big(T_{\rm R,max} + \frac{T_{10}}{e^{ T_{10} / T_{\rm bg} } - 1 } \big) ) }~. 
\end{equation}

In reality, however, molecular clouds are highly inhomogeneous. As long as the line is optically thick, one can only observe emission that originates from the cloud surface up to a thickness of $\tau\approx 1$ where the density is lower and thus $T_{\rm exc}$ is lower. Therefore, Eq.~\eqref{eq:Texc_obs} only measures the excitation temperature at the cloud surface (and hence the subscript~``s''). In the upper panels of Fig.~\ref{fig:NCO_vs_temp_Z}, the blue solid lines show $T_{\rm exc,s}$ as a function of $N_{\rm CO}$ for $Z^{\prime} = 3, 1, 0.3, \text{and } 0.1$ from left to right. Note that we only show $T_{\rm exc,s}$ in the optically thick regime ($\tau_{10}>1$) where Eq.~\eqref{eq:Texc_obs} is applicable. At~$Z^\prime = 0.1$, $\tau_{10} < 1$ everywhere and thus $T_{\rm exc,s}$ is not shown. Comparing with the $n_{\rm CO}$-weighted l.o.s.\ average excitation temperature:
\begin{equation}\label{eq:Texc_los_ave}
    \overline{T}_{\rm exc} \equiv \frac{\int n_{\rm CO} T_{\rm exc} \,\mathrm{d}z}{ \int n_{\rm CO} \,\mathrm{d}z }~,
\end{equation}
as shown by the orange dashed lines, we see that  $T_{\rm exc,s}$ is indeed slightly lower than $\overline{T}_{\rm exc}$ though the effect is mild (less than a factor of two). Finally, the green dotted lines are the $n_{\rm CO}$-weighted l.o.s.\ average kinetic temperature:
\begin{equation}\label{eq:Tkin_los_ave}
    \overline{T}_{\rm kin} \equiv \frac{\int n_{\rm CO} T_{\rm kin} \,\mathrm{d}z}{ \int n_{\rm CO} \,\mathrm{d}z }~.
\end{equation}
$\overline{T}_{\rm exc}$ gradually increases and approaches $\overline{T}_{\rm kin}$ as $N_{\rm CO}$ increases. Note that $\overline{T}_{\rm exc} < \overline{T}_{\rm kin}$ at almost all $N_{\rm CO}$. Therefore, the fact that 
$\overline{T}_{\rm exc}$ (and thus $T_{\rm exc,s}$)
increases with $N_{\rm CO}$ is mainly driven by the increase of average density instead of a temperature effect as $\overline{T}_{\rm kin}$ is nearly constant.
This also explains why $W_{10}$ does not scale as $\ln(N_{\rm CO})$ in the optically thick regime ($N_{\rm CO}  \gtrsim 10^{16}~{\rm cm^{-2}}$), as the theory of curve of growth would predict. Instead, $W_{10}$ increases more rapidly with $N_{\rm CO}$ (see Fig.~\ref{fig:nco_vs_tau_wco}), as $T_{\rm exc,s}$ provides an extra contribution to the growth of $W_{10}$. 

Similarly, the lower panels of Fig.~\ref{fig:NCO_vs_temp_Z} show the line ratio ($R_{21}$) in blue solid lines as a function of $N_{\rm CO}$ for $Z^{\prime} = 3, 1, 0.3, \text{and } 0.1$ from left to right. The ``optically thin'' line ratio
\begin{equation}\label{eq:Rthin}
    R_{\rm 21,thin} \equiv \frac{W_{\rm 21,thin}}{ W_{\rm 10,thin} } = 2.4 \frac{\overline{f}_2}{\overline{f}_1}
\end{equation}
(from Eqs.~\eqref{eq:W10_thin} and~\eqref{eq:W21_thin}) is shown by orange dashed lines while the ``LTE optically thin'' line ratio 
\begin{equation}\label{eq:RthinLTE}
    R_{\rm 21,LTE}(T_{\rm kin}) \equiv 2.4 \frac{ \overline{f}_2(T_{\rm kin}) }{ \overline{f}_1(T_{\rm kin}) }
\end{equation}
is shown by green dotted lines.

In the optically thin regime, $R_{21} = R_{\rm 21,thin}$ by definition. At the lowest $N_{\rm CO}$, the excitation is set by the CMB temperature and therefore $R_{21} = R_{\rm 21,thin} = 2.4 f_2(2.73) / f_1(2.73) = 0.069$. As $N_{\rm CO}$ increases, the average density also increases which gradually drives $R_{\rm 21,thin}$ towards $R_{\rm 21,LTE}$, similar to the behavior of $r_{21}$ shown in Fig.~\ref{fig:nH_vs_x_T_R21}. The maximum of $R_{\rm 21,thin}$ is ${\sim}2.5$ and is insensitive to $Z^\prime$ (as is the case for $r_{21}$). Note that the maximum theoretical line ratio is $R_{\rm 21,LTE}(\infty) = 4$. In addition, at lower $Z^\prime$, $R_{\rm 21}$ rises up at a lower $N_{\rm CO}$ because of the density effect as shown in Fig.~\ref{fig:nco_vs_tau_wco}. At high $N_{\rm CO}$, $R_{\rm 21}$ falls slightly below $R_{\rm 21,thin}$ by about a factor of two due to the optical depth effect: as both lines become optically thick, the observed $R_{\rm 21}$ originates from the cloud surfaces with thickness of $\tau\approx 1$, where the density is lower and the level population is more subthermal. However, at $Z^\prime = 0.1$, both lines are only marginally optically thick and thus $R_{\rm 21}$ becomes closer to its optically thin limit $R_{\rm 21,thin}$, reaching a maximum of ${\sim}1.5$, (about 50\% larger than the higher $Z^\prime$ cases).

\section{Observational implications}\label{sec:obs_impl}

\subsection{Spatially Resolved \texorpdfstring{$X_{\rm CO}$}{XCO} and \texorpdfstring{$R_{21}$}{R21}}\label{sec:2pcXCO}

\begin{figure*}
	\centering
	\includegraphics[width=0.99\linewidth]{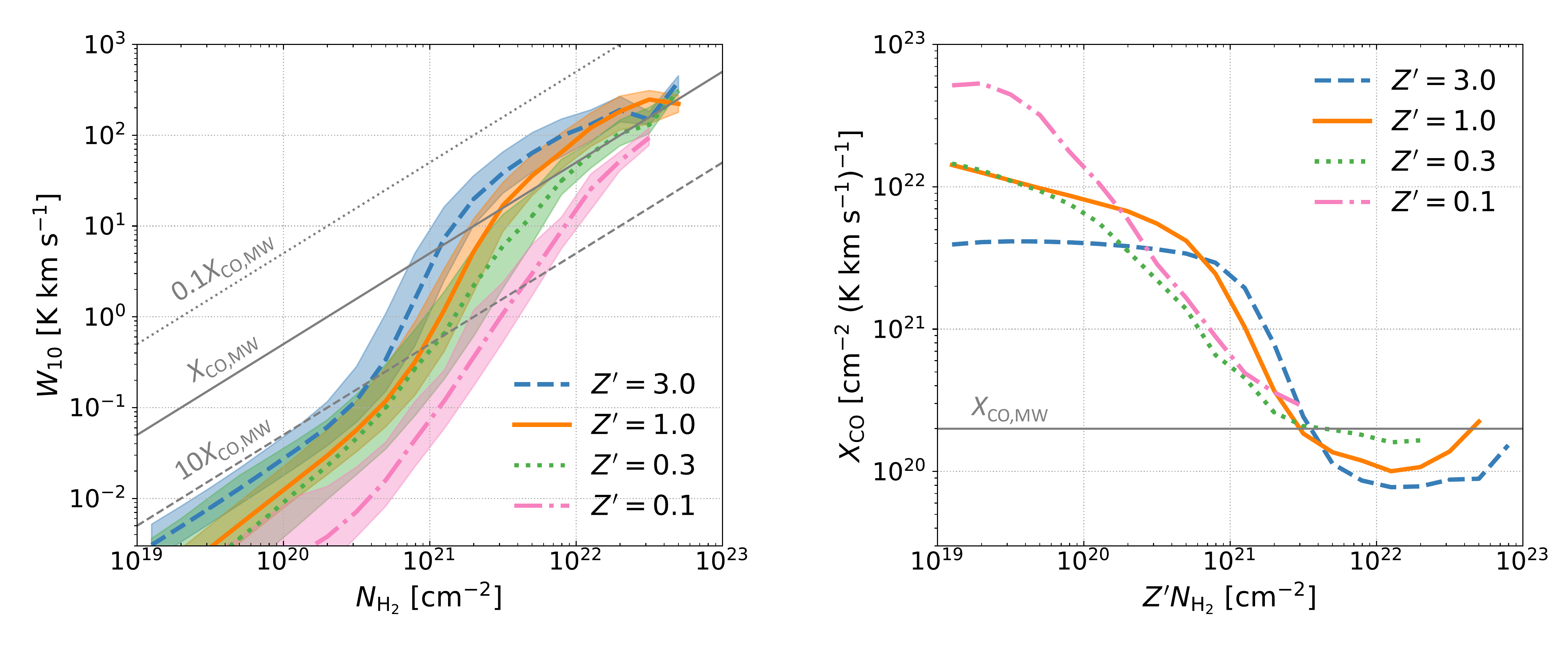}
	\caption{%
		\textit{Left}: \coone\ line intensity as a function of H$_2$ column density ($N_{\rm H_2}$). The lines represent the median values while the shaded areas enclose the 25th and 75th percentiles. The dotted, solid, and dashed lines in grey indicate $X_{\rm CO} / X_{\rm CO,MW} = 0.1, 1, \text{and } 10$, respectively, The variation of $X_{\rm CO}$ is mainly driven by the variation of CO abundance. \textit{Right}: $X_{\rm CO}$ as a function of $Z^\prime N_{\rm H_2}$. The scatter is not shown for clarity. $X_{\rm CO}$ drops to the Milky Way value ($X_{\rm CO,MW}$, horizontal grey line) at $Z^\prime N_{\rm H_2} \sim 3\times 10^{21}\ {\rm cm^{-3}}$ (visual extinction $A_V \sim 3$) where dust shielding becomes effective.
	}
	\label{fig:NH2_vs_WCO_XCO_Z_scatter_final}
\end{figure*}

Armed with the detailed background in Section~\ref{sec:mol_physics},
we are now in a good position to study the small-scale distributions of $X_{\rm CO}$ and $R_{21}$. The left panel of Fig.~\ref{fig:NH2_vs_WCO_XCO_Z_scatter_final} shows $W_{10}$ as a function of $N_{\rm H_2}$ at different $Z^\prime$. The dotted, solid, and dashed lines in grey indicate $X_{\rm CO} / X_{\rm CO,MW} = 0.1, 1, \text{and } 10$, respectively. Recall from Fig.~\ref{fig:nco_vs_tau_wco} that $W_{10} < 10~{\rm K~km~s^{-1}}$ (or $N_{\rm CO} < 10^{16}~{\rm cm^{-3}}$)  is the optically thin regime where $W_{10} \propto N_{\rm CO}$. It is also in this regime that $X_{\rm CO}$ varies the most, which corresponds to the formation of CO (see Fig.~\ref{fig:summary_chem}). At $W_{10} > 10~{\rm K~km~s^{-1}}$, the line becomes optically thick and $W_{10}$ flattens, except for the $Z^\prime = 0.1$ case where the line remains optically thin. In this optically thick regime, $X_{\rm CO}$ only varies mildly. Therefore, the variation of $X_{\rm CO}$ mostly occurs in the optically thin regime and is driven by the variation of CO abundance. Note that even at $Z^\prime = 0.1$, $X_{\rm CO}$ eventually reaches $X_{\rm CO,MW}$ at the highest $N_{\rm H_2}$. In fact, if we plot $X_{\rm CO}$ as a function of $Z^\prime N_{\rm H_2}$, as in the right panel of Fig.~\ref{fig:NH2_vs_WCO_XCO_Z_scatter_final}, we see that for all $Z^\prime$, $X_{\rm CO}$ drops to $X_{\rm CO,MW}$ (the horizontal grey line) at $Z^\prime N_{\rm H_2} \sim 3\times 10^{21}~{\rm cm^{-3}}$, which corresponds to a visual extinction $A_V \sim 3$ and is where dust shielding becomes effective.
This is qualitatively consistent with the cloud-scale simulations in \citet{2011MNRAS.412.1686S}.



\begin{figure*}
	\centering
	\includegraphics[width=1.\linewidth]{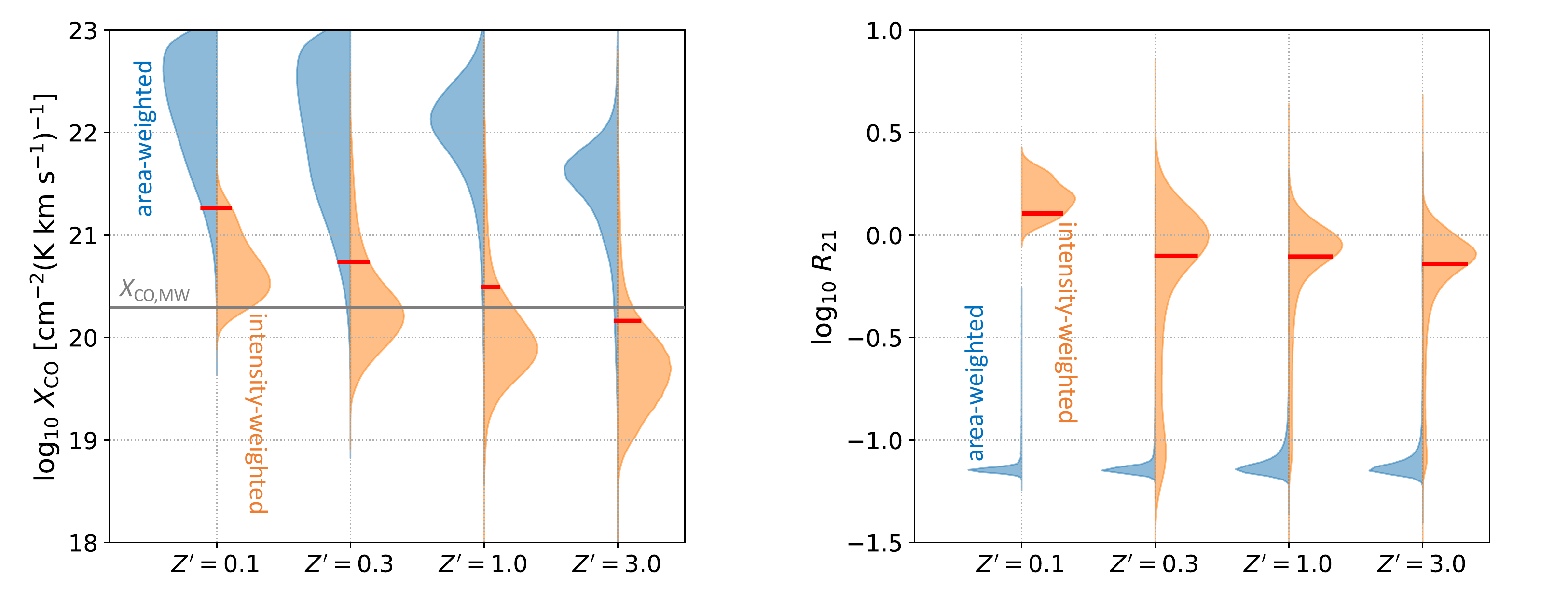}
	\caption{%
		Probability density functions (PDFs) of $X_{\rm CO}$ (left panel) and $R_{21}$ (right panel) at different $Z^{\prime}$. The area-weighted PDFs are shown in blue while the $W_{10}$-weighted PDFs are shown in orange. The $W_{10}$-weighted average of  $X_{\rm CO}$ (Eq.~\eqref{eq:XCO_aveW}) and $R_{21}$ (Eq.~\eqref{eq:R21_aveW}) are indicated by the red bars. The Milky-Way value of $X_{\rm CO}$ is shown by the horizontal grey line. Most cloud area is filled by diffuse gas with high $X_{\rm CO}$ and low $R_{21}$, while most \coone\ emission originates from dense gas with low $X_{\rm CO}$ and high $R_{21}$.
	}
	\label{fig:violin_XCO_R21_final}
\end{figure*}

Fig.~\ref{fig:violin_XCO_R21_final} shows the probability density functions (PDFs) of $X_{\rm CO}$ (left panel) and $R_{21}$ (right panel) at different $Z^{\prime}$. The area-weighted PDFs are shown in blue while the $W_{10}$-weighted PDFs are shown in orange.
Most cloud area is filled by diffuse gas with high $X_{\rm CO}$ and low $R_{21}$, while most \coone\ emission originates from dense gas with low $X_{\rm CO}$ and high $R_{21}$.

As $X_{\rm CO}$ is a distribution rather than a constant, if we want choose a statistical quantity to represent the distribution, a natural choice would be the $W_{10}$-weighted average (as shown by the red horizontal bars):
\begin{equation}\label{eq:XCO_aveW}
	\langle X_{\rm CO} \rangle_W \equiv \frac{\int X_{\rm CO} W_{10} \,\mathrm{d}a }{ \int W_{10} \,\mathrm{d}a}  
	= \frac{\sum_i X_{{\rm CO},i} W_{10,i}}{\sum_i W_{10,i}}~,
\end{equation}
where the summation is over the $512^2$ beams. On the other hand, the conversion factor in a coarse beam of size~$l_{\rm b}$ can be expressed as
\begin{equation}\label{eq:XCOpix}
    X_{\rm CO}(l_{\rm b}) = \frac{N_{\rm H_2}(l_{\rm b})}{W_{10}(l_{\rm b})}
    = \frac{\sum_i N_{{\rm H_2},i}}{\sum_i W_{10,i}}
    = \frac{\sum_i X_{{\rm CO},i} W_{10,i}}{\sum_i W_{10,i}}~,
\end{equation}
where the summation is over all sub-beams within a coarse beam. Therefore, $\langle X_{\rm CO} \rangle_W$ is also the global conversion factor for the entire (1~kpc)$^2$ area shown in Fig.~\ref{fig:intro_global_Z_XCO}.
Similarly, the global $W_{10}$-weighted average of $R_{21}$ (red horizontal bars) is
\begin{equation}\label{eq:R21_aveW}
	\langle R_{21} \rangle_W \equiv \frac{\int R_{21} W_{10} \,\mathrm{d}a }{ \int W_{10} \,\mathrm{d}a} 
	= \frac{\sum_i R_{21,i} W_{10,i} }{\sum_i W_{10,i}}~.
\end{equation}
Since the line ratio in a coarse beam of size~($l_{\rm b}$) can be written as
\begin{equation}\label{eq:R21pix}
    R_{21}(l_{\rm b}) = \frac{W_{21}(l_{\rm b})}{W_{10}(l_{\rm b})}
    = \frac{\sum_i W_{21,i}}{\sum_i W_{10,i}}
    = \frac{\sum_i R_{21,i} W_{10,i}}{\sum_i W_{10,i}}~,
\end{equation}
where the summation is over all sub-beams within a coarse beam, $\langle R_{21} \rangle_W$ is also the global line ratio for the entire (1~kpc)$^2$ area.

\subsection{Scale Dependence}\label{sec:scale}

\begin{figure*}
	\centering
	\includegraphics[width=0.99\linewidth]{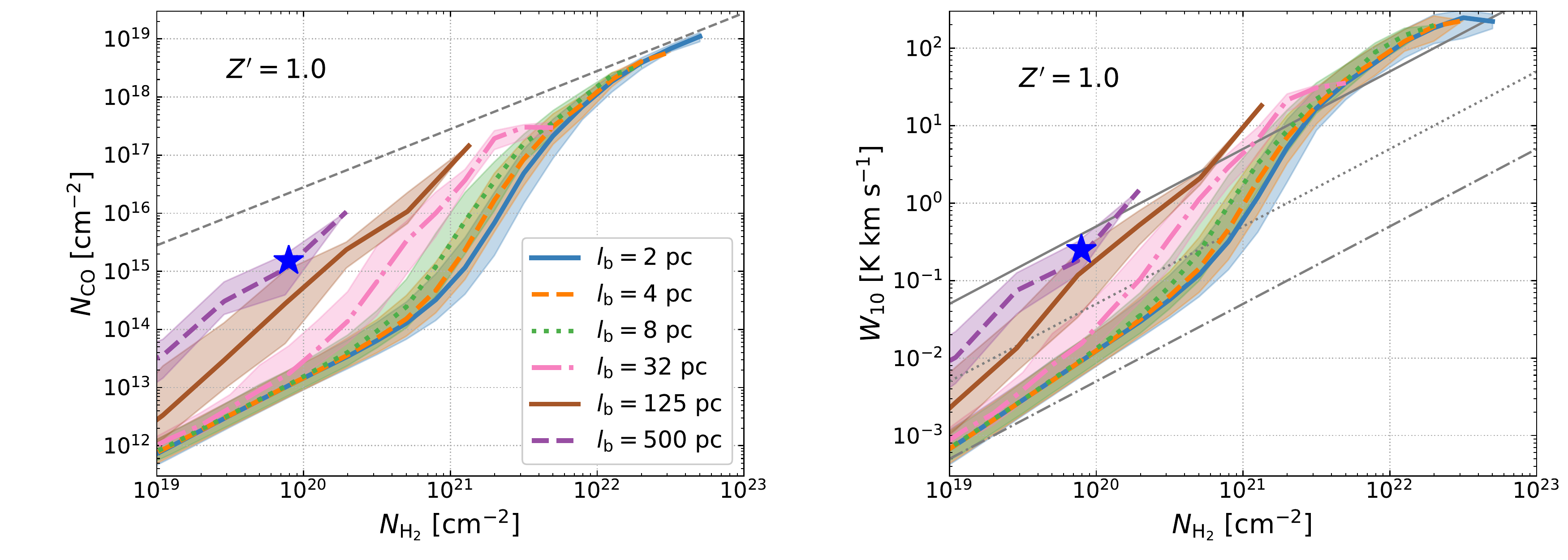}
	\caption{%
		\textit{Left}: Relationships between $N_{\rm H_2}$ and $N_{\rm CO}$ at different beam sizes ($l_{\rm b}$) for $Z^{\prime} = 3, 1, 0.3, \text{and } 0.1$ (from top to bottom). The lines represent the median values while the shaded areas enclose the 25th and 75th percentiles. The dashed grey lines indicate $N_{\rm CO} = 2 N_{\rm H_2} x_{\rm C,0}$ where hydrogen and carbon are fully in the forms of H$_2$ and CO, respectively. \textit{Right}: Same as the left panels, but with $W_{10}$ in the $y$-axis. The solid, dotted and dash-dotted lines in grey indicate $X_{\rm CO} / X_{\rm CO,MW} = 1, 10, \text{and } 100$, respectively, where $X_{\rm CO,MW} = 2\times 10^{20}~{\rm cm^{-2}\ (K~km~s^{-1})^{-1}}$ is the canonical CO-to-H$_2$ conversion factor. The blue star symbols indicate the time-averaged values with $l_{\rm b} = 1$~kpc.
	}
	\label{fig:NH2_vs_NCO_WCO10_pix}
\end{figure*}

The left panels in Fig.~\ref{fig:NH2_vs_NCO_WCO10_pix} show the relationships between $N_{\rm H_2}$ and $N_{\rm CO}$ at different beam sizes for $Z^{\prime} = 1$.
The dashed grey lines indicate $N_{\rm CO} = 2 N_{\rm H_2} x_{\rm C,0}$, which is the upper limit of $N_{\rm CO}$ when carbon is completely in form of CO.

\begin{figure*}
	\centering
	\includegraphics[width=0.99\linewidth]{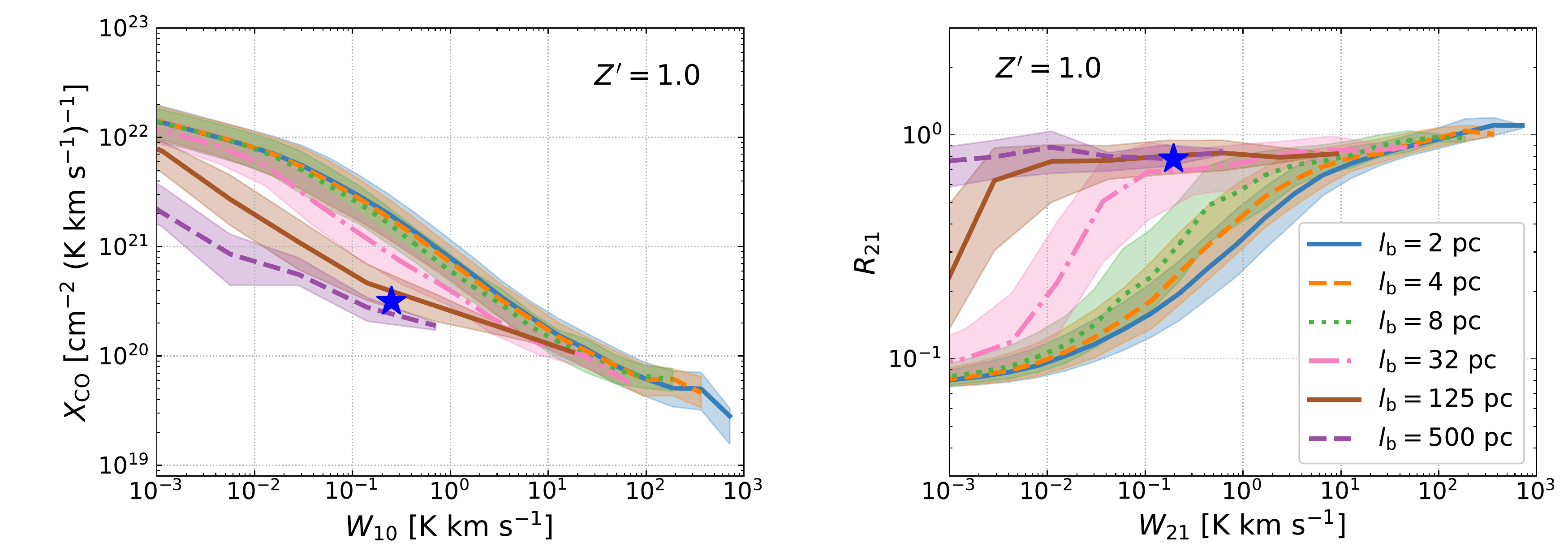}
	\caption{%
		\textit{Left}: $X_{\rm CO}$ as a function of $W_{10}$ with different beam sizes ($l_{\rm b}$) for $Z^{\prime} = 3, 1, 0.3, \text{and } 0.1$ (from top to bottom). The lines represent the median values while the shaded areas enclose the 25th and 75th percentiles. \textit{Right}: Same as the left panels, but showing the line ratio ($R_{21} \equiv W_{21} / W_{10}$) as a function of $W_{21}$. The blue star symbols indicate the time-averaged values with $l_{\rm b} = 1$~kpc.
	}
	\label{fig:XCO_R21_pix}
\end{figure*}

\begin{figure*}
	\centering
	\includegraphics[width=0.99\linewidth]{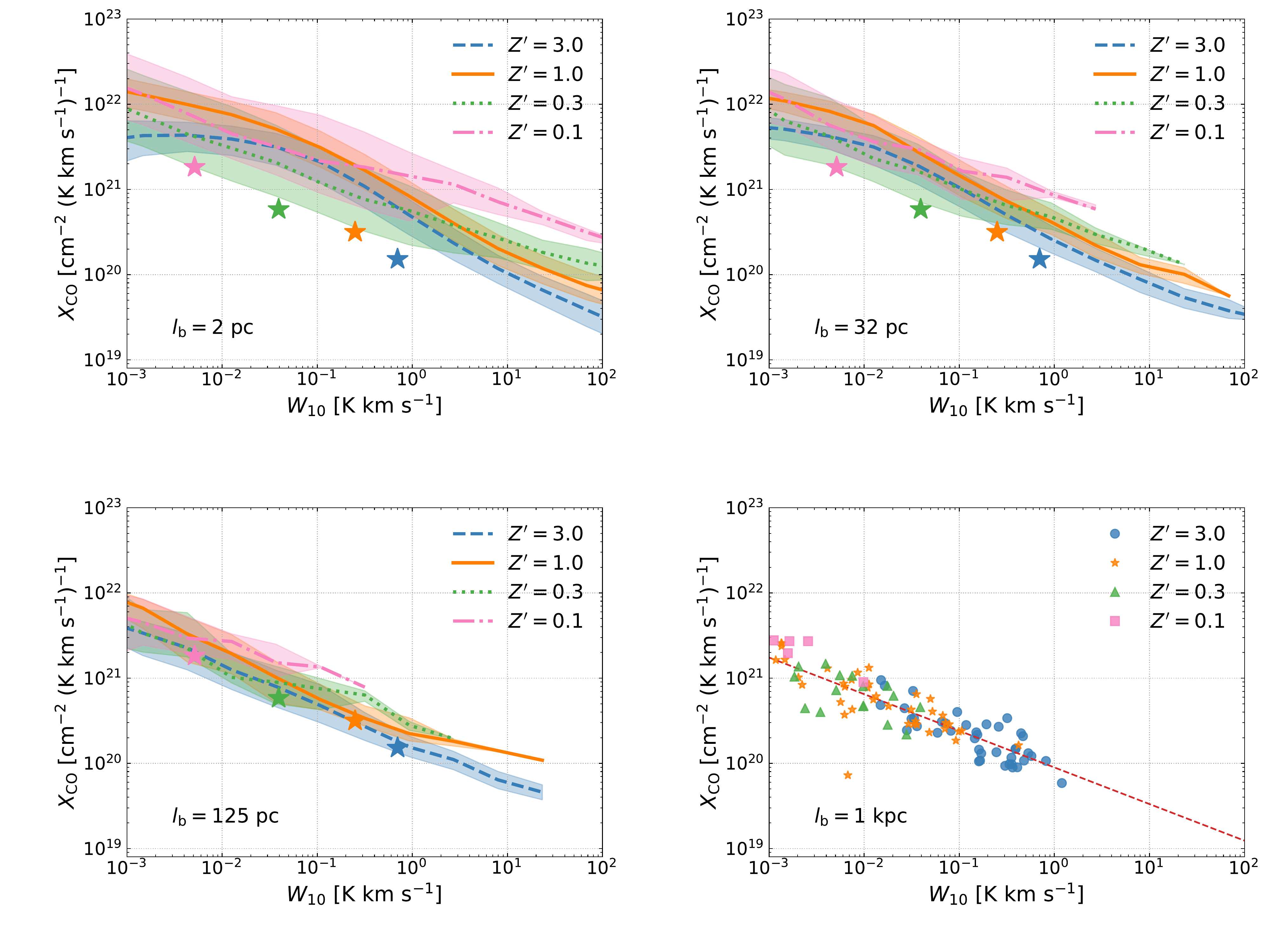}
	\caption{%
		$X_{\rm CO}$ as a function of $W_{10}$ for different $Z^\prime$ (colored lines) and different $l_{\rm b}$ (panels). The lines represent the median values while the shaded areas enclose the 25th and 75th percentiles. The time-averaged global values are shown by the star symbols. The bottom right panel is shown with a scatter plot due to limited data points at $l_{\rm b} = 1$~kpc.
		The $W_{10}$ -- $X_{\rm CO}$ relation at $l_{\rm b} = 1$~kpc can be well-described by a single power law (red dashed line, Eq.~(\ref{eq:fit2})) 
		for all metallicities.
		This figure can be used to more accurately infer H$_2$ mass observationally.
	}
	\label{fig:W10_vs_XCO_pix_2by2}
\end{figure*}

\begin{figure*}
	\centering
	\includegraphics[width=0.99\linewidth]{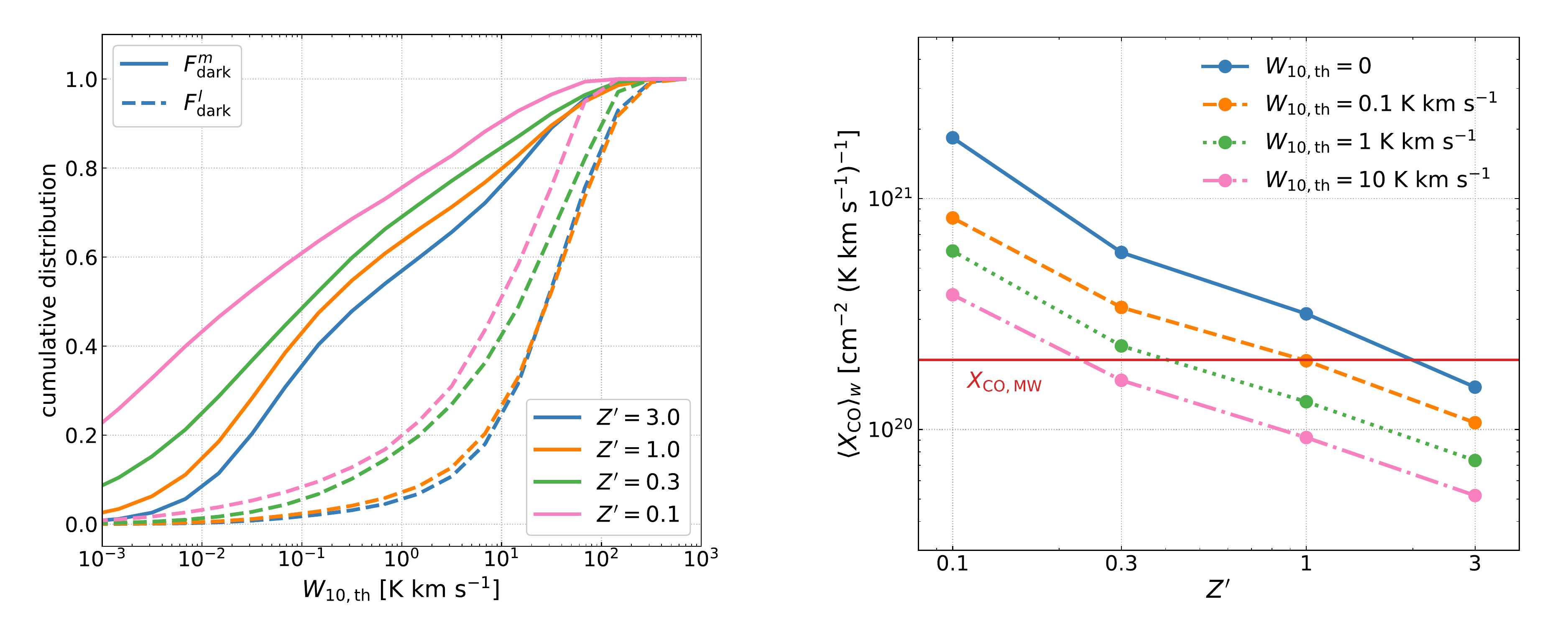}
	\caption{
		\textit{Left}: CO-dark H$_2$ mass fraction (solid) and CO-dark light fraction (dashed) as a function of detection threshold $W_{\rm 10, th}$.
		\textit{Right}: Intensity-weighted average $X_{\rm CO}$ as a function of $Z^\prime$ with different detection thresholds. The horizontal line shows the Milky Way value $X_{\rm CO,MW}$.
		A higher detection threshold leads to a lower $X_{\rm CO,MW}$ as only the brightest gas with low $X_{\rm CO}$ can be detected.
	} 
	\label{fig:Z_vs_XCO_Wth_final}
\end{figure*}

On large scales, a coarse beam contains a distribution of $N_{\rm H_2}$ and $N_{\rm CO}$, and the CO-to-H$_2$ ratio can be expressed as
\begin{equation}\label{eq:rCOpix}
	\gamma (l_{\rm b})
		= \frac{ N_{\rm CO} (l_{\rm b})}{N_{\rm H_2} (l_{\rm b})}
		= \frac{\sum_i N_{{\rm CO},i} }{\sum_i N_{{\rm H_2},i} }
		= \frac{\sum_i \gamma_i N_{{\rm H_2},i} }{\sum_i N_{{\rm H_2},i} }~,
\end{equation}
where the summation is over all sub-beams within a coarse beam. Namely, the coarse-beam $\gamma$
is the $N_{\rm H_2}$-weighted average of the small-scale~$\gamma$ over the sub-beam distribution and the dense, CO-rich gas (where $\gamma$ is high) has a dominant contribution to the coarse-beam~$\gamma$.
Since
$\gamma$ increases monotonically with $N_{\rm H_2}$, the coarse-beam~$\gamma$
increases with $l_{\rm b}$ at a given $N_{\rm H_2}$. In addition, the variation of $\gamma$ decreases as $l_{\rm b}$ increases because of beam averaging/\linebreak[0]{}smoothing.

Qualitatively, the $N_{\rm H_2}{-}W_{10}$ relation, which we show in the right panels of Fig.~\ref{fig:NH2_vs_NCO_WCO10_pix}, is very similar to the $N_{\rm H_2}{-}N_{\rm CO}$ relation.
The solid, dotted and dash-dotted lines in grey indicate $X_{\rm CO} / X_{\rm CO,MW} = 1, 10, \text{and } 100$, respectively. 
Recall that the conversion factor in a coarse beam is the $W_{10}$-weighted average of the small-scale $X_{\rm CO}$ over the sub-beam distribution. If $X_{\rm CO}$ is constant within a coarse beam (even if $N_{\rm H_2}$ and $W_{10}$ are not constant), the coarse-beam conversion factor is unchanged. However, if there is a sub-beam distribution of $X_{\rm CO}$, Eq.~\eqref{eq:XCOpix} implies that the dense, CO-bright gas (where $X_{\rm CO}$ is low) has a dominant contribution to the coarse-beam $X_{\rm CO}$.
Since $W_{10}$ is a super-linear function of $N_{\rm H_2}$ (i.e., $X_{\rm CO}$ decreases as $N_{\rm H_2}$ increases), the coarse-beam conversion factor decreases with $l_{\rm b}$ at a given $N_{\rm H_2}$. In addition, the variation of $X_{\rm CO}$ decreases as $l_{\rm b}$ increases because of the smoothing/\linebreak[0]{}averaging.

As $l_{\rm b}$ decreases, the $N_{\rm H_2}{-}W_{10}$ relationship gradually converges. Convergence occurs when $l_{\rm b}$ is small enough such that $X_{\rm CO}$ is constant within each beam. To achieve this, the transition regions where $X_{\rm CO}$ varies most have to be spatially resolved, which is increasingly challenging as $Z^\prime$ decreases since the CO-bright clouds becomes smaller.

To make connections with observables, we show $X_{\rm CO}$ as a function of $W_{10}$ in the left panels of Fig.~\ref{fig:XCO_R21_pix}. 
The blue star symbol indicates the time-averaged global ($l_{\rm b} = 1$~kpc) value. On small scales, $X_{\rm CO}$ varies by orders of magnitude. The observed variation depends on the detection threshold. If the threshold intensity $W_{\rm 10,th}$ is high (i.e., low sensitivity), the variation of $X_{\rm CO}$ is reduced as only the brightest regions with low $X_{\rm CO}$ can be detected while the diffuse gas becomes ``CO-dark.''%
\footnote{%
	Note that we adopt an observational definition of ``CO-dark.'' Namely, we distinguish between ``CO-dark'' gas and ``high-$X_{\rm CO}$'' gas. The former solely reflects the detectability controlled by the detection threshold $W_{\rm 10,th}$ and is, in principle, independent of $X_{\rm CO}$. Gas with low $X_{\rm CO}$ can be CO-dark if $W_{\rm 10,th}$ is high. Conversely, Gas with high $X_{\rm CO}$ can also be CO-bright if $W_{\rm 10,th}$ is low enough to detect it.}
On the other hand, if we adopt a much lower $W_{\rm 10,th}$ (i.e., better sensitivity),
$X_{\rm CO}$ can vary by more than an order of magnitude as both low- and high-$X_{\rm CO}$ regions are detected. In this case, adopting a constant $X_{\rm CO}$ (which is typically done in observations) implies that $N_{\rm H_2}$ would be substantially underestimated in diffuse gas and overestimated in dense gas. In other words, the gradient of $N_{\rm H_2}$ would be overestimated.

As $l_{\rm b}$ decreases, $X_{\rm CO}$ becomes increasingly uniform due to beam averaging
and eventually approaches the global value. For example, at $l_{\rm b} = 125$~pc,
$X_{\rm CO}$ only varies by a factor of two above $W_{\rm 10} = 1~{\rm K~km~s^{-1}}$.

The scale dependence of $R_{21}$ can be understood in a similar way, as shown in the right panels of Fig.~\ref{fig:XCO_R21_pix}. $R_{21}$ in a coarse beam is the $W_{10}$-weighted average of the small-scale $R_{21}$ over the sub-beam distribution and the dense, thermalized gas (where $R_{21}$ is high) has a dominant contribution to the coarse-beam $R_{21}$.
Since $R_{21}$ increases monotonically with $W_{\rm 10}$, the coarse-beam $R_{21}$ increases with $l_{\rm b}$. The variation of $R_{21}$ decreases as $l_{\rm b}$ increases because of the smoothing/\linebreak[0]{}averaging
and eventually $R_{21}$ approaches the global average at $l_{\rm b} = 1$~kpc (blue star symbol). 

Figs.~\ref{fig:NH2_vs_NCO_WCO10_pix} and~\ref{fig:XCO_R21_pix} focus exclusively on solar metallicity. In Fig.~\ref{fig:W10_vs_XCO_pix_2by2}, we show $X_{\rm CO}$ as a function of $W_{10}$ for different $Z^\prime$ (colored lines) and different $l_{\rm b}$ (panels). The time-averaged global ($l_{\rm b} = 1$~kpc) values are shown in stars. On small scales ($l_{\rm b} = 2$~pc), $X_{\rm CO}$ is a decreasing function of $W_{10}$  with a secondary dependence on $Z^\prime$ most prominent at $W_{10} > 1~{\rm K~km~s^{-1}}$. As $l_{\rm b}$ increases, the $W_{10} {-} X_{\rm CO}$ relation is shifted downward and rightward. 
Moreover, the secondary dependence on $Z^\prime$ is gradually reduced for a given $W_{10}$. 
At $l_{\rm b} = 1$~kpc, a single power-law relation
holds for all $Z^\prime$, with different $Z^\prime$ populating different ranges of $W_{10}$.
We provide the best-fit formula shown as the red dashed line:
\begin{equation}\label{eq:fit2}
X_{\rm CO} = 8.95\times10^{19} \Big(\frac{W_{10}}{{\rm K~km~s^{-1}}} \Big)^{-0.43} {\rm cm^{-2}~(K~km~s^{-1})^{-1}},	
\end{equation}
with a correlation coefficient of 0.90.

Fig.~\ref{fig:W10_vs_XCO_pix_2by2} summarizes the key result of this paper and can be used to more accurately infer the H$_2$ mass from \coone\ observations. It highlights the fact that $X_{\rm CO}$ depends not only on $Z^\prime$, but also on $W_{10}$ and $l_{\rm b}$. Different observations probe different regimes of the parameter space, leading to discrepancies when projecting on the $Z^\prime {-} X_{\rm CO}$ plane. For example, observations using the dust-based method or the inverse KS~method typically have kpc-scale beam sizes within which the diffuse and dense gas are averaged over, while high-resolution CO observations (the virial method) are naturally biased towards the dense gas  as the diffuse gas might be below the detection threshold.

\subsection{CO-dark \texorpdfstring{H$_2$}{H2} Fraction}

The left panel of Fig.~\ref{fig:Z_vs_XCO_Wth_final} shows the cumulative distributions of $W_{10}$ weighted by $N_{\rm H_2}$ (solid) and $W_{10}$ (dashed), respectively. Most \coone\ emission is coming from  $1 < W_{10} < 100\ {\rm K~km~s^{-1}}$, while the H$_2$ mass is distributed in a lower and more widespread range $10^{-2} < W_{10} < 10\ {\rm K~km~s^{-1}}$. This difference makes \coone\ an imperfect tracer of H$_2$. The $N_{\rm H_2}$-weighted cumulative distribution can also be interpreted as the CO-dark H$_2$ \textit{mass} fraction
\begin{equation}
	F^m_{\rm dark} \equiv \frac{M_{\rm H_2}(W_{\rm 10} < W_{\rm 10,th})}{M_{\rm H_2}}
\end{equation}
as a function of detection threshold ($W_{\rm 10,th}$). Adopting $W_{\rm 10,th} = 1\ {\rm K~km~s^{-1}}$, the CO-dark H$_2$ mass fraction is about 55\% to 75\% from $Z^\prime = 3$ to~$0.1$, consistent with \citetalias{HSvD21} where we adopted an $N_{\rm CO}$-equivalent threshold. More than half of the H$_2$ mass is hidden in the undetected diffuse gas. Similarly, the $W_{10}$-weighted cumulative distribution can be interpreted as the CO-dark \textit{light} fraction
\begin{equation}
    F^l_{\rm dark} \equiv \frac{L_{10}(W_{10} < W_{\rm 10,th})}{L_{10}}
\end{equation}
as a function of detection threshold ($W_{\rm 10,th}$). For $W_{\rm 10,th} = 1~{\rm K~km~s^{-1}}$, the CO-dark light fraction is about 5\% to 20\% from $Z^\prime = 3$ to~$0.1$, i.e., most of the \coone\ emission can be detected.

Correspondingly, the intensity-weighted average conversion factor over the detectable beams (i.e., $W_{10} \geq W_{\rm 10,th}$)
as a function of $Z^\prime$ is shown in the right panel of Fig.~\ref{fig:Z_vs_XCO_Wth_final}. The Milky Way value $X_{\rm CO,MW}$ is shown by the horizontal red line. As the CO-bright H$_2$ mass fraction is $1 - F^m_{\rm dark}$ and the CO-bright light fraction is $1 - F^l_{\rm dark}$, the intensity-weight conversion factor above a detection threshold $W_{\rm 10,th}$ follows
\begin{align}\label{eq:COdark}
	\langle X_{\rm CO} (W_{\rm 10} \geq W_{\rm 10,th}) \rangle_W
	&= \frac{1 - F^m_{\rm dark}}{1 - F^l_{\rm dark}} \langle X_{\rm CO} \rangle_W \nonumber\\
	&\approx (1 - F^m_{\rm dark}) \langle X_{\rm CO} \rangle_W~,
\end{align}
where the approximation holds when most of the \coone\ emission can be detected (i.e., $1 - F^l_{\rm dark} \approx 1$). For example, if we adopt $W_{\rm 10,th} = 1\ {\rm K\ km\ s^{-1}}$, then  $F^m_{\rm dark} \approx 60\%$, $F^l_{\rm dark} \approx 5\%$, and thus $\langle X_{\rm CO} (W_{\rm 10} \geq 1\ {\rm K\ km\ s^{-1}}) \rangle_W  \approx 0.42 \langle X_{\rm CO} \rangle_W$. In general, a higher detection threshold leads to a lower value as only the brightest gas with low $X_{\rm CO}$ can be detected.


Table~\ref{tab:global} summarizes several global intensity-weighted average quantities.

\begin{deluxetable}{cccccc}
	\tablecaption{Global intensity-weighted average quantities.
		\label{tab:global}
	}
	\tablewidth{0pt}
	\tablehead{
		\colhead{$ Z^{\prime} $ }  &
		\colhead{$\langle X_{\rm CO}                 \rangle_W $ }  &
		\colhead{$\langle R_{21}                     \rangle_W $ }  &
		\colhead{$\langle \tau_{10}                  \rangle_W $ }  &
		\colhead{$\langle \tau_{21}                  \rangle_W $ }  &
		\colhead{$\langle \overline{T}_{\rm exc,s} \rangle_W $ }  
	}
	\decimalcolnumbers
	\startdata
	3   &  $1.52\times 10^{20}$	   &  0.71	 &  39.0   &   99.3    &  11.4     \\
	1   &  $3.17\times 10^{20}$		 &  0.78   &  15.5   &   42.8    &  12.3   	 \\
	0.3 &  $5.84\times 10^{20}$	   &  0.79   &  3.38   &   10.8    &  10.6     \\
	0.1 &  $1.83\times 10^{21}$		 &  1.29	 &  0.31   &   1.38    &  --    \\
	\enddata
	\tablecomments{
		(1) Normalized metallicity.
		(2) CO-to-H$_2$ conversion factor in units of ${\rm cm^{-2}\ (K~km~s^{-1})^{-1}}$.
		(3) Line ratio of \cotwo\ to \coone.
		(4) Optical depth of \coone.
		(5) Optical depth of \cotwo\ (weighted by $W_{21}$ instead of $W_{10}$).
		(6) Cloud surface excitation temperature of \coone\ in units of~K assuming optically thick conditions (see Eq. \ref{eq:Texc_obs}). It is absent in the $Z^\prime = 0.1$ case as \coone\ is always optically thin even in the densest gas.
	}
\end{deluxetable}

\section{Summary}\label{sec:summary}

We have studied the CO-to-H$_2$ conversion factor and  the line ratio of \cotwo\ to \coone\ in high-resolution (${\sim}0.2$~pc) hydrodynamical simulations 
of a multiphase ISM across a wide range of metallicity ($0.1 \leq Z^\prime \leq 3$) in \citetalias{HSvD21}. We use the radiative transfer code {\sc RADMC-3D} to post-process CO emission over a 400~Myr time period such that a large number of clouds at different evolutionary stages are properly sampled. We interpolate the particle data onto an adaptive mesh (the code is publicly available\footnote{\href{https://github.com/huchiayu/ParticleGridMapper.jl}{https://github.com/huchiayu/ParticleGridMapper.jl}}) with a minimum cell size of ${\sim}0.2$~pc to ensure that all small-scale CO emission is properly captured in our radiative transfer calculations, which is particularly important at low metallicities. Our main findings can be summarized as follows:

\begin{enumerate}
\item The kpc-scale $X_{\rm CO}$ at low $Z^\prime$ can be overestimated  either by assuming steady-state chemistry 
(e.g., when converting fine structure metal lines and CO lines to an H$_2$ mass via PDR models) or by assuming the star-forming gas is fully molecular (e.g., when using the inverse KS~method to convert an SFR to an H$_2$ mass) (see Fig.~\ref{fig:intro_global_Z_XCO}).
Our fiducial, time-dependent model can be described by Eq.~(\ref{eq:fit1})
on 1 kpc scale. 

\item 
Instead of a single-variable function of $Z^\prime$,
$X_{\rm CO}$ is a multivariate function of three observables: $Z^\prime$, $W_{10}$, and $l_{\rm b}$ (Fig.~\ref{fig:W10_vs_XCO_pix_2by2}). 
Observations with different beam sizes and detection thresholds are sensitive to different parts of the parameter space.
Comparing $X_{\rm CO}$ between different observations
on the $Z^\prime{-}X_{\rm CO}$ plane
can therefore be misleading,
and a fair comparison should take all three variables into account.

\item
On large scales,
the metallicity dependence on the $W_{10}{-}X_{\rm CO}$ plane is gradually reduced
as a result of the sub-beam gas distribution.
At $l_{\rm b} = 1$~kpc, a single power-law relation (Eq.~(\ref{eq:fit2}))
holds for all $Z^\prime$, 
with different $Z^\prime$ populating different ranges of $W_{10}$ (Fig.~\ref{fig:W10_vs_XCO_pix_2by2}).

\item On parsec scales, $X_{\rm CO}$ varies by orders of magnitude from place to place.
The variation of $X_{\rm CO}$ occurs primarily in the optically thin regime, 
driven by the variation of CO abundance in the transition from atomic carbon to CO. The pc-scale $X_{\rm CO}$ can be well-characterized by a single parameter $Z^\prime N_{\rm H_2}$, and it drops to the Milky Way value once dust shielding becomes effective (Fig.~\ref{fig:NH2_vs_WCO_XCO_Z_scatter_final}).

\item The line intensities are proxies of column or volume densities and thus can be used to infer $X_{\rm CO}$ and $R_{21}$. On pc-scales, $X_{\rm CO}$ negatively correlates with $W_{10}$ while $R_{21}$ positively correlates with $W_{21}$ (Fig.~\ref{fig:XCO_R21_pix}). 
The large-scale $X_{\rm CO}$ and $R_{21}$ are results of beam averaging,
and they are naturally biased towards values in dense gas with low $X_{\rm CO}$ and high $R_{21}$ (Eqs.~\eqref{eq:XCOpix} and~\eqref{eq:R21pix}).

\item Most cloud area is filled by diffuse gas with high $X_{\rm CO}$ and low $R_{21}$, while most CO emission originates from dense gas with low $X_{\rm CO}$ and high $R_{21}$ (Fig.~\ref{fig:violin_XCO_R21_final}). Adopting a constant $X_{\rm CO}$ strongly over- \mbox{(under-)}\linebreak[0]{}estimates H$_2$ in dense (diffuse) gas.

\item Both \coone\ and \cotwo\ are thermalized at higher densities as $Z^\prime$ decreases (Fig.~\ref{fig:nH_vs_x_T_R21}) because (i) H$_2$ exists at densities higher than the critical densities of the lines, and (ii) the CO abundances and hence the optical depth are lower which leads to less radiation trapping.

\item The l.o.s.\ average gas density increases with $N_{\rm CO}$. In addition, as CO only exists at the densest core at low $Z^\prime$, the corresponding l.o.s.\ average gas density is higher at a given $N_{\rm CO}$. This effect counters the higher thermalization densities  such that the CO levels are more efficiently excited at low $Z^\prime$ (Fig.~\ref{fig:nco_vs_tau_wco}).

\item The CO lines become increasingly optically thin at lower $Z^\prime$. This is not only because of the lower CO column densities but also because the levels are more efficiently excited, which leads to a smaller optical depth at a given $N_{\rm CO}$ (Fig.~\ref{fig:nco_vs_tau_wco}).
In the optically thick regime, the line intensities increase faster than $\ln N_{\rm CO}$ (as the curve of growth theory would predict) because the excitation temperature also increases with $N_{\rm CO}$ and gradually approaches the kinetic temperature due to the increase in the l.o.s.\ average density.

\item The excitation temperature derived from the peak radiation temperature (Eq.~\eqref{eq:Texc_obs}) is sub-thermal
(Fig.~\ref{fig:NCO_vs_temp_Z} and Table~\ref{tab:global}). At $Z^\prime = 0.1$, Eq.~\eqref{eq:Texc_obs} is no longer applicable 
as the optically thick assumption breaks down. In addition, $R_{21}$ increases as the lines become increasingly optically thin.

\item The intensity-weighted average $X_{\rm CO}$ is controlled by the CO-dark H$_2$ mass fraction and the CO-dark light fraction (Fig.~\ref{fig:Z_vs_XCO_Wth_final} and Eq.~\eqref{eq:COdark}), and it decreases with the detection threshold $W_{\rm 10,th}$ as only the brightest gas with low $X_{\rm CO}$ can be detected.

\end{enumerate}

To sum up,
$X_{\rm CO}$ is a multivariate function of metallicity, line intensity, and beam size. 
Although we did not find a simple parametrization for this function,
our Fig.~\ref{fig:W10_vs_XCO_pix_2by2} can be used by observers to more accurately infer the H$_2$ mass in galaxies with similar conditions of the solar neighborhood.
It is still unclear
to what extent this function is applicable in different environments (e.g., high-redshift galaxies, dwarf galaxies, etc.) where the large-scale physical properties (e.g., gas surface density, turbulence, etc.) are very different.
This will be investigated in followup studies.

\section*{Acknowledgments}
We thank the anonymous referee for the constructive comments that improved our manuscript.
C.Y.H. acknowledges support from the DFG via German-Israel Project Cooperation grant STE1869/2-1 GE625/17-1.
A.S. thanks the Center for Computational Astrophysics (CCA) of the Flatiron Institute,
and the Mathematics and Physical Science (MPS) division of the Simons Foundation for support.
All simulations were run on the Raven, Cobra and Draco supercomputers at the Max Planck Computing and Data Facility (MPCDF).

\bibliography{literatur}{}
\bibliographystyle{aasjournal}

\appendix

\section{Scale dependence of \texorpdfstring{$\langle X_{\rm CO} \rangle_W$}{<XCO>w}}

\begin{figure*}
	\centering
	\includegraphics[width=0.99\linewidth]{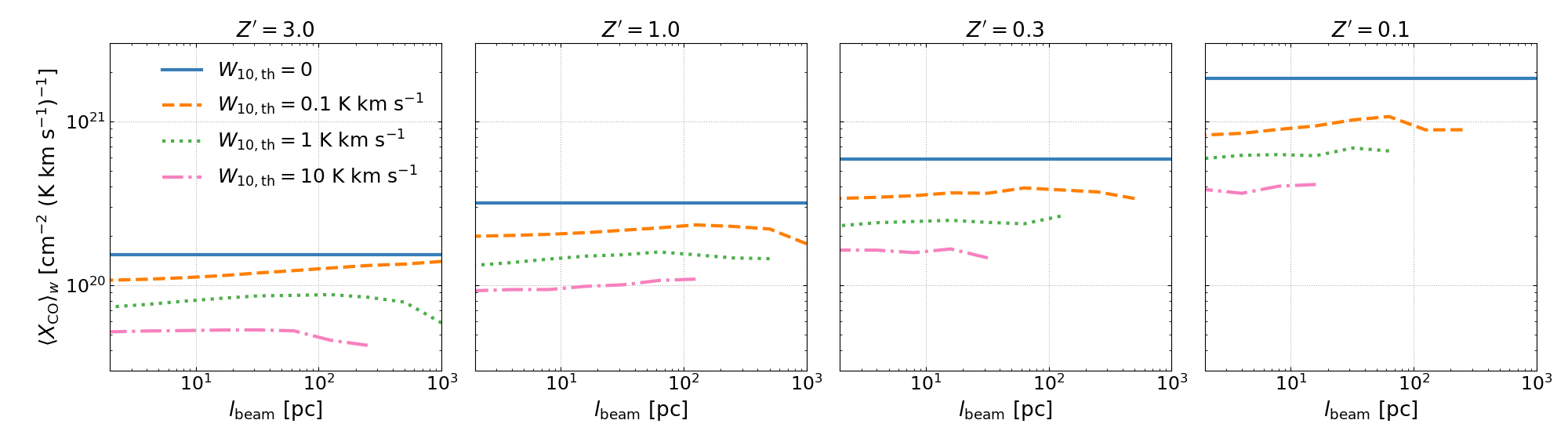}
	\caption{
		Global intensity weighted average CO-to-H$_2$ conversion factor as a function of beam size with different detection thresholds $W_{\rm 10,th}$ at $Z^\prime = 3, 1, 0.3, \text{and } 0.1$ from left to right.
	}
	\label{fig:lpix_vs_XCO_Z}
\end{figure*}

\begin{figure*}
	\centering
	\includegraphics[width=0.99\linewidth]{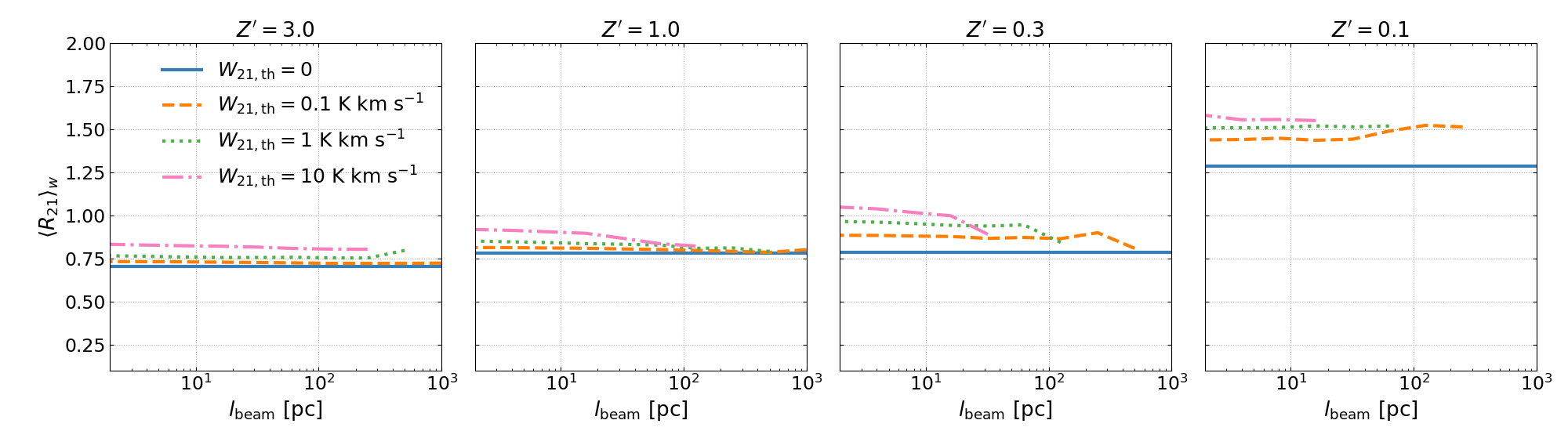}
	\caption{
		Global intensity-weighted average line ratio as a function of beam size with different detection thresholds $W_{\rm 21,th}$ at $Z^\prime = 3, 1, 0.3, \text{and } 0.1$ from left to right.
	}
	\label{fig:lpix_vs_R21_Z}
\end{figure*}

In the case of $W_{\rm 10,th} = 0$ (i.e., infinite sensitivity),
the global intensity-weighted average $\langle X_{\rm CO} \rangle_W$
is by construction independent of $l_{\rm b}$
as beam averaging is equivalent to the $W_{10}$-weighted average over the sub-beam distribution
(see Eqs.~\eqref{eq:XCO_aveW} and~\eqref{eq:XCOpix}).
However,
if we consider a nonzero $W_{\rm 10,th}$,
$\langle X_{\rm CO} \rangle_W$ is only averaged over
the CO-bright beams where $W_{10} > W_{\rm 10,th}$
and thus does depend on $l_{\rm b}$,
as shown in Fig.~\ref{fig:lpix_vs_XCO_Z}.
At a given $l_{\rm b}$,
$\langle X_{\rm CO} \rangle_W$
decreases inversely with $W_{\rm 10,th}$
as 
the faint, high $X_{\rm CO}$ gas becomes CO-dark
and 
only the bright, low $X_{\rm CO}$ gas can be detected.
With $W_{\rm 10,th} = 0$,
$\langle X_{\rm CO} \rangle_W$ is indeed (by construction) independent of $l_{\rm b}$.
With a finite $W_{\rm 10,th}$,
there is a very mild $l_{\rm b}$-dependence
that arises from two competing effects.
On one hand,
a coarse beam averages over CO-bright and CO-dark gas.
As CO emission predominately originates from the CO-bright gas,
the coarse-beam intensity is likely to exceed the detection threshold.
Namely,
the entire coarse beam becomes CO-bright,
which effectively includes the fine-grained CO-dark gas within the beam to the global average,
leading to an increase in 
$\langle X_{\rm CO} \rangle_W$ 
as CO-dark gas tends to be of high $X_{\rm CO}$.
On the other hand,
when the beam size is so large that
the majority of the area is filled with CO-dark gas,
the dilution effect can be strong enough such that the entire coarse beam becomes CO-dark,
which effectively excludes the fine-grained CO-bright gas within the beam from the global average.
Therefore,
as $l_{\rm b}$ increases,
$\langle X_{\rm CO} \rangle_W$
first increase at small $l_{\rm b}$ and then decreases at large $l_{\rm b}$.
The effect is very mild,
within 30\% from $l_{\rm b} = 2$~pc to $l_{\rm b} = 1$~kpc.
In short,
$\langle X_{\rm CO} \rangle_W$
mainly depends on the detection threshold (as shown in Fig.~\ref{fig:Z_vs_XCO_Wth_final})
and is nearly scale-independent.

Similarly,
Fig.~\ref{fig:lpix_vs_R21_Z}
demonstrates the scale-dependence of the global average line ratio
$\langle R_{21} \rangle_W$
(see Eq.~\eqref{eq:R21_aveW})
with different detection thresholds $W_{\rm 21,th}$.
$\langle R_{21} \rangle_W$
is nearly scale-independent
and it increases with $W_{\rm 21,th}$
as $R_{21}$ tends to be high in CO-bright gas.
However,
the variation is very weak, less than 20\% in all cases.

\section{Scale dependence of the CO-dark Gas}

\begin{figure*}
	\centering
	\includegraphics[width=0.99\linewidth]{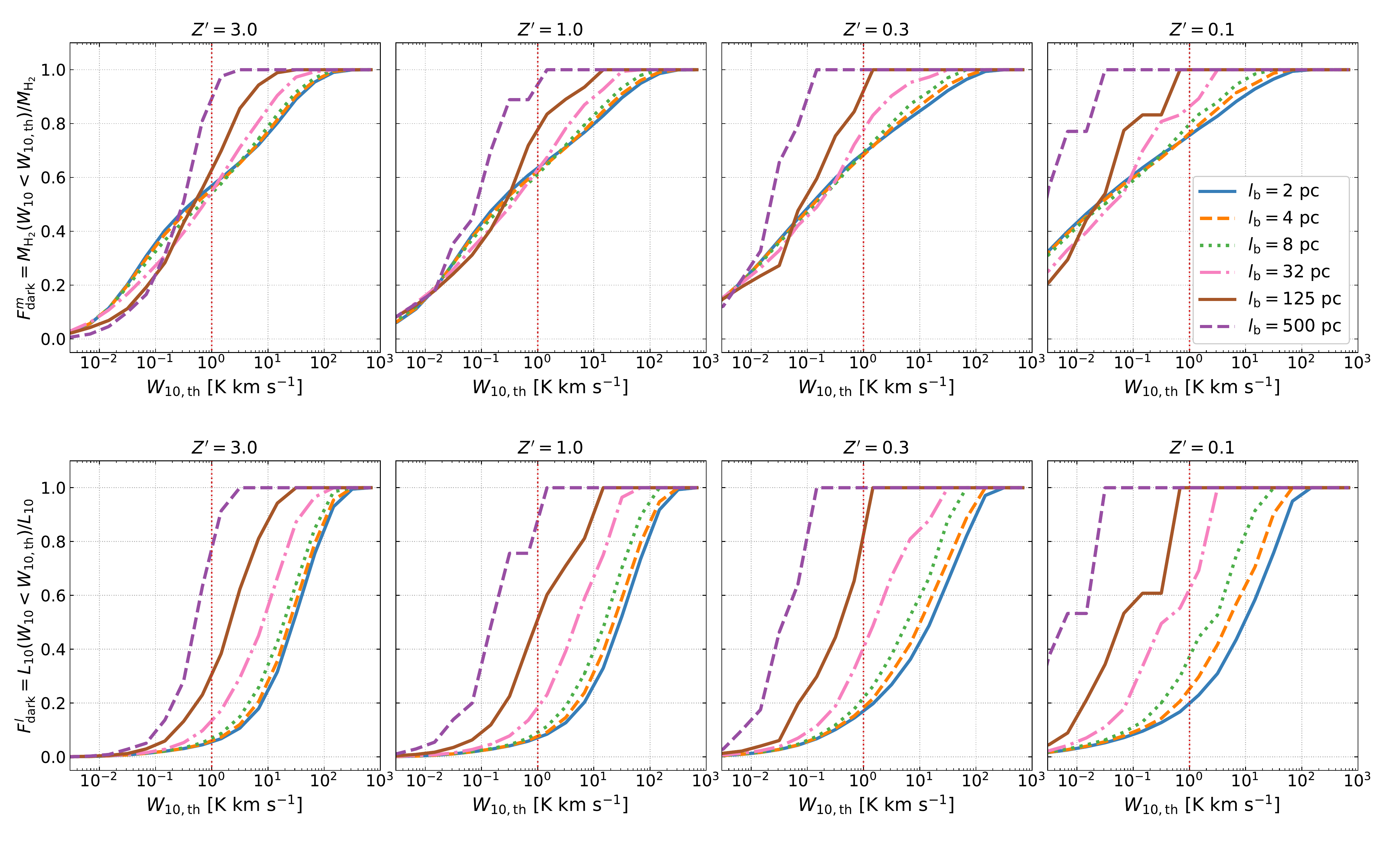}
	\caption{
	    CO-dark H$_2$ mass fraction (upper) and CO-dark light fraction (lower) as a function of detection threshold $W_{\rm 10, th}$ at different beam sizes $l_{\rm b}$
		for $Z^\prime = 3, 1, 0.3, \text{and } 0.1$ from left to right.
	}
	\label{fig:hist_COdark_pix}
\end{figure*}

For a given $W_{\rm 10,th}$,
$\langle X_{\rm CO} \rangle_W$ 
only accounts for H$_2$ in the CO-bright beams,
while a significant amount of H$_2$ may be hidden in the CO-dark beams.
Fig.~\ref{fig:hist_COdark_pix}
shows the CO-dark H$_2$ mass fraction ($F^m_{\rm dark}$, upper row) and CO-dark light fraction ($F^l_{\rm dark}$, lower row) as a function of detection threshold $W_{\rm 10, th}$ at different beam sizes $l_{\rm b}$ for $Z^\prime = 3, 1, 0.3, \text{and } 0.1$ from left to right.
As $l_{\rm b}$ increases,
the CDF becomes narrower due to beam smoothing
and gradually approaches a step function.
Therefore,
$F^m_{\rm dark}$
may increase or decrease as $l_{\rm b}$ increases,
depending on the detection threshold.
For $W_{\rm 10,th} = 1\ {\rm K\ km\ s^{-1}}$,
$F^m_{\rm dark}$ remains almost constant at small $l_{\rm b}$
where the distribution is converged.
At $l_{\rm b} \gtrsim 32$~pc,
$F^m_{\rm dark}$ increases with $l_{\rm b}$ 
as the beam averaging 
makes the fine-grained CO-bright gas undetectable.
Similarly,
$F^l_{\rm dark}$ remains constant at small $l_{\rm b}$
and increase with $l_{\rm b}$ at $l_{\rm b} \gtrsim 32$~pc.
This provides another explanation of why
$\langle X_{\rm CO} \rangle_W$ is insensitive to $l_{\rm b}$,
as both $F^m_{\rm dark}$ and $F^l_{\rm dark}$ begin to increase at $l_{\rm b} \gtrsim 32$~pc and they cancel out (see Eq.~\eqref{eq:COdark}).

\section{\coone\ spectrum}\label{app:spectra}

\begin{figure}
	\centering
	\includegraphics[trim=0cm 0cm 1cm 0cm,clip, width=1\linewidth]{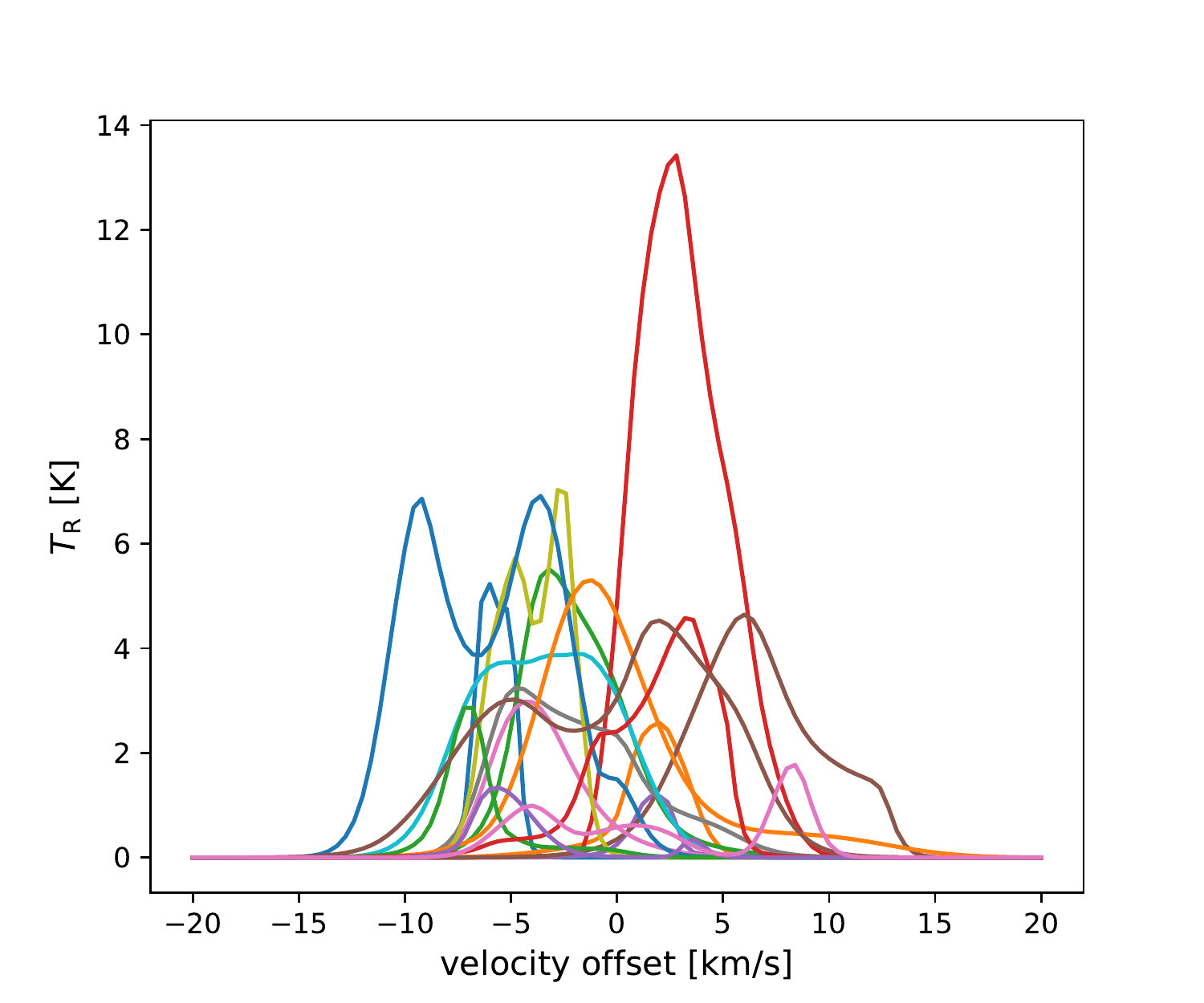}
	\caption{
	\coone\ spectra randomly drawn from the $512 \times 512$ pixels in the $Z^\prime = $~1 run at $t = 420$~Myr. Gas motion leads to a variety of spectrum profiles, including Gaussian-like, skewed, saturated, or even double-peaked.
	}
	\label{fig:spectra}
\end{figure}

Fig.~\ref{fig:spectra} shows 2500 \coone\ spectra randomly drawn from the $512 \times 512$ pixels in the $Z^\prime = $~1 run at $t = 420$~Myr.
Gas motion leads to a variety of spectrum profiles, including Gaussian-like, skewed, saturated, or even double-peaked. Our adopted spectrum resolution ($0.4~{\rm km~s^{-1}}$) is sufficient to resolve the typical line width of the spectra, and our spectrum coverage ($40~{\rm km~s^{-1}}$) is large enough to include all the \coone\ emission.

\section{Interpolation Scheme}\label{app:interp}

\begin{figure*}
	\centering
	\includegraphics[width=1\linewidth]{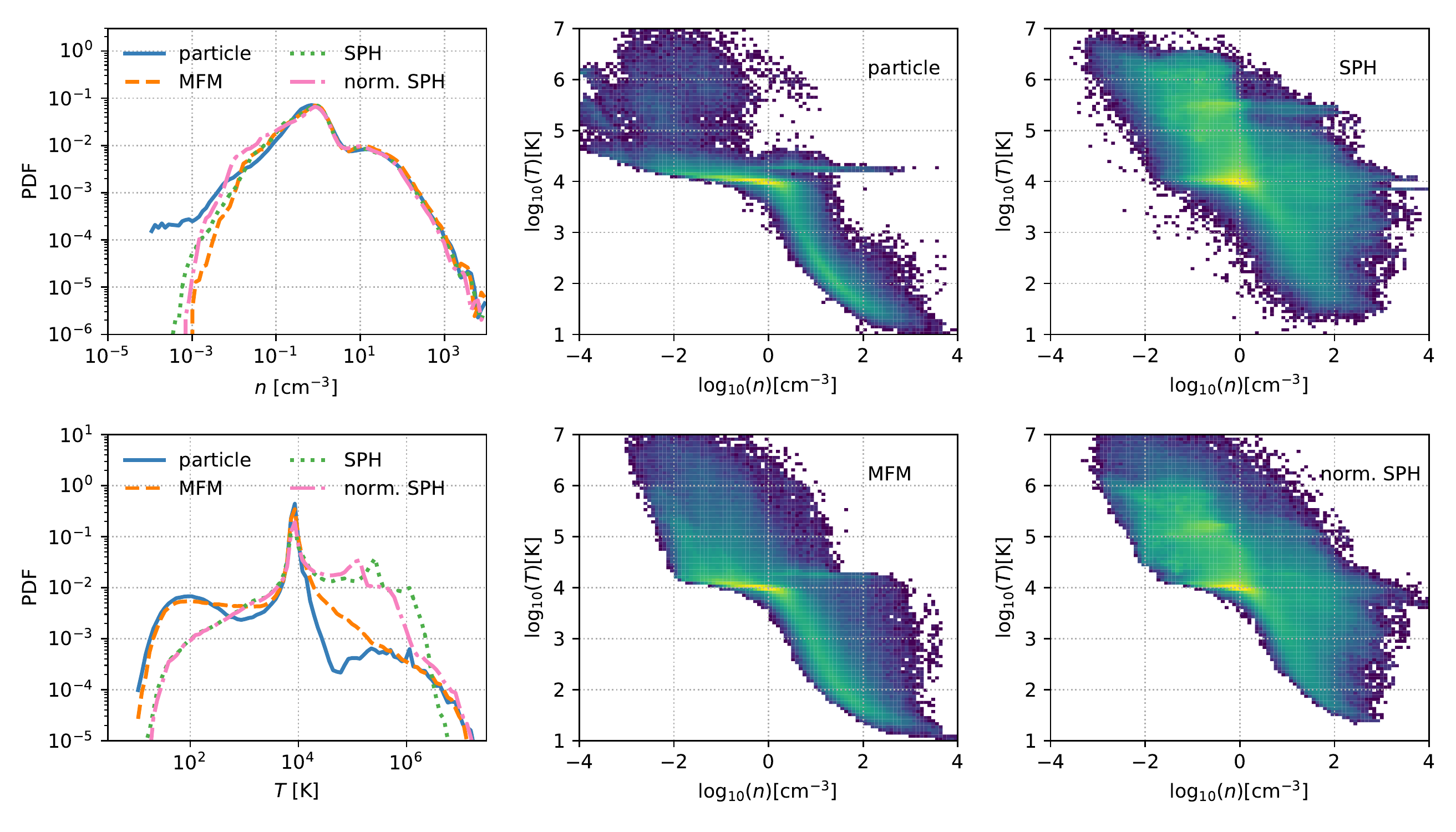}
	\caption{
		\textit{Left panels}:
		Mass-weighted PDFs of 
		hydrogen number density ($n$, upper panel) and temperature ($T$, lower panel)
		for the particle data,
		the standard SPH scheme (Eq.~\eqref{eq:SPH}),
		the normalized SPH scheme (Eq.~\eqref{eq:nSPH}),
		and the MFM scheme (Eq.~\eqref{eq:MFM}).
		\textit{Middle and right panels}:
		``phase diagram'' (heatmap of $n$ vs.~$T$)
		for the particle data and
		the three interpolation schemes.
		The MFM scheme faithfully reproduces the distributions of $n$ and $T$,
		while the two SPH schemes significantly over-estimate $T$.
	}
	\label{fig:compare}
\end{figure*}

In this appendix,
we compare three different interpolation schemes that map Lagrangian (particle) data onto a mesh.
For a given particle distribution,
where a particle $i$ 
carries information of
mass ($m_i$), gas density ($\rho_i$), smoothing length ($h_i$), and a scalar field ($A_i$)
at location $x_i$.
The standard SPH scheme
interpolate the scalar field
at location $x_c$ as
\begin{equation}\label{eq:SPH}
    \bar{A}_{\rm SPH}(x_c) =  \sum_{j} \frac{m_j}{\rho_j} A_j K(|x_j - x_c|, h_j)~,
\end{equation}
where the summation is over the ``scatter'' neighboring particles $j$.
This scheme has the drawback
that a constant scalar field would be interpolated as non-constant,
as $\sum_{j} \frac{m_j}{\rho_j} K(|x_j - x_c|, h_j)$, in general, does not sum up to unity.
This can be remedied by the normalized SPH scheme:
\begin{equation}\label{eq:nSPH}
\bar{A}_{\rm nSPH}(x_c) =  
\frac{ \sum_{j} \frac{m_j}{\rho_j} A_j K(|x_j - x_c|, h_j) }{ \sum_{j} \frac{m_j}{\rho_j} K(|x_j - x_c|, h_j) }~,
\end{equation}
which factors out the non-unity term.
Alternatively,
one can use the MFM scheme: 
\begin{equation}\label{eq:MFM}
\bar{A}_{\rm MFM}(x_c) = \frac{ \sum_{j} A_j K(|x_j - x_c|, h_j) }{ \sum_{j} K(|x_j - x_c|, h_j) }~.
\end{equation}
Similar to the normalized SPH scheme,
the MFM scheme interpolates a constant field as constant.

In Fig.~\ref{fig:compare},
the left panels of
show the mass-weighted PDFs of 
hydrogen number density ($n$, upper panel) and temperature ($T$, lower panel)
for the particle data and
the three interpolation schemes.
The middle and right panels
show 
the mass-weighted ``phase diagram'' (heatmap of $n$ vs.~$T$)
for the particle data and
the three interpolation schemes.

While all schemes reproduce the actual density distribution from the particle data,
both the standard SPH scheme and normalized SPH scheme 
significantly over-produce the hot gas ($10^4 < T < 10^6$~K)
and under-produce the cold gas ($10 < T < 10^3$~K).
This is caused by the volume factor $m_j/\rho_j$
which put a strong weighting on the diffuse gas (low $\rho_j$).
For density interpolation,
the $\rho_j$ dependence cancels out.
For temperature interpolation,
as the diffuse gas typically has a high $T$,
the contribution from the hot gas is enormous,
and thus the result is generally biased high.

In contrast,
the MFM scheme 
depends merely on particle configuration (which defines $h_j$)
and is independent of $m_j$ and $\rho_j$.
This means that the diffuse gas does not have the same strong weighting as in the SPH schemes,
and 
thus the temperature distribution is more faithfully reproduced.

\end{document}